\documentclass[journal]{IEEEtran}
\usepackage{graphicx}
\usepackage{enumitem} 
\usepackage[table,xcdraw]{xcolor}
\usepackage{multirow}
\usepackage{bm}
\usepackage{colortbl}
\usepackage{algorithm}
\usepackage{algpseudocode}
\usepackage{tikz}

\usepackage{url}
\usepackage{adjustbox}
\usepackage{booktabs}
\usepackage{pgfplots}
\pgfplotsset{compat=1.18}
\usepackage{pgfplotstable}
\usepackage{siunitx}
\usepackage{array}
\NewDocumentCommand{\todo}{O{red} +m}{\begingroup\color{#1}#2\endgroup}
\newcommand*\circled[1]{\tikz[baseline=(char.base)]{
            \node[shape=circle,draw,inner sep=0.8pt] (char) {#1};}}

\ifCLASSINFOpdf
\else
\fi

\usepackage{amsmath}
\begin{document}
%
\title{FDM: A \textbf{F}ramework for \textbf{D}ecision-making to build ML-based \textbf{M}alware detection systems}

%
%
%

\author{\IEEEauthorblockN{Tadiwa Vhito, Jakapan Suaboot*, Warodom Werapun, Norrathep Rattanavipanon
\thanks{${}^*$Corresponding author}}\\
\IEEEauthorblockA{\textit{College of Computing}\\
\textit{Prince of Songkla University}\\
Phuket, Thailand \\
\{s6530621001, jakapan.su, wwarodom, norrathep.r\}@phuket.psu.ac.th}

}
\maketitle

\begin{abstract}
Selecting appropriate machine learning (ML) configurations for malware detection is a complex, multi-criteria problem: model choice, feature engineering strategy, and update mechanism must jointly satisfy operational constraints that vary substantially across deployment contexts. This paper proposes a Framework for Decision-making to build ML-based Malware detection systems (FDM), a structured methodology that formalises this selection process through the Weighted Configuration Compatibility Score (WCCS), a multi-criteria scoring function that maps five quantifiable operational parameters (platform constraint, resource budget, response latency, update frequency, and detection sensitivity) to ranked recommendations across nine configuration dimensions. To validate the framework, four experiments were conducted on three datasets (a private Windows API call dataset covering eight malware families, the public Malimg image benchmark of 9,339 samples from 25 families, and a public Android static API dataset of 15,036 applications). Key results include: (i)~XGBoost achieved the best accuracy-to-resource ratio in binary classification (97.46\% test accuracy, $<$70\,MB RAM), outperforming LSTM and BiLSTM which consumed up to 2.8\,GB; (ii)~in multi-class classification, classical models (XGBoost 79.03\%) outperformed recurrent deep models (BiLSTM 72.27\%), reversing the binary ranking; (iii)~class-incremental learning with EfficientNetB0 maintained 99.13\% accuracy with only 0.65 percentage-point accuracy degradation across eleven incremental steps; (iv)~transfer learning reduced training time by an average of $2.14\times$ on image-based malware data with no significant accuracy cost; and (v)~autoencoder pre-processing yielded a $14\times$ training speedup at a cost of only 0.86\,pp accuracy. These findings collectively confirm that the optimal ML configuration is context-dependent, validating the FDM's core premise and demonstrating its practical utility for cybersecurity practitioners.
	 
\end{abstract}


%
\IEEEpeerreviewmaketitle

\section{Introduction}
The rapid growth of artificial intelligence (AI) technology has caused an unprecedented surge in malware attacks, presenting a critical challenge in modern cybersecurity. This rapid evolution is evidenced by the discovery of more than 30 million new malware strains and potentially unwanted programs in just the first half of 2026 as reported by the AV-TEST Institute\footnote{https://portal.av-atlas.org/malware}. According to the World Economic Forum\footnote{https://www.weforum.org/stories/2023/06/asia-pacific-region-the-new-ground-zero-cybercrime/}, the average annual cost of cybercrime is
expected to increase to more than \$23 trillion by 2027. Therefore, an urgent advancement in defensive capabilities is required.

Malware detection presents significant challenges due to the sophisticated evasion techniques employed by malware creators. These evasion techniques can generally be categorized into three groups \cite{geng2024survey}: (i) transformation-based methods (such as metamorphism, polymorphism \cite{Polymo}, dead code insertion and instruction replacement or reordering \cite{Evade}), (ii) concealment-based tactics (including delay execution and code obfuscation), and (iii) attack-based techniques (such as exploitation attacks \cite{zhang2022semantics}). The use of anti-sandboxing techniques further complicates dynamic analysis-based detection methods, as malware can be designed to postpone or evade malicious actions when it identifies a sandbox environment \cite{cyberbit19}. Certain evasion methods, notably polymorphism and instruction replacement or reordering, allow attackers to generate malware variants that maintain the same functionalities while displaying different signatures, thus successfully evading detection, particularly by signature-based techniques. Furthermore, zero-day attacks, which exploit undiscovered vulnerabilities in software or operating systems, render prevention efforts almost futile \cite{KAKAREKA20141}. Consequently, there is a continuous demand for researchers to develop innovative malware detection methods (e.g. \cite{malimg,9800057,info13120563}).

Malware detection has evolved from signature-based methods using predefined patterns, e.g., YARA rules \cite{yaramahdi2024detection}, to identify known threats. Although effective for simple malware and offering low false alarms, signature-based approaches can easily be circumvented through the evasion techniques mentioned above to modify malware signatures and thwart the detection system. Hence, machine learning (ML) has revolutionized malware detection, becoming a primary tool to identify malicious behavior instead of a static signature. Deep learning, in particular, excels at revealing complex patterns that classical (non-deep) learning methods cannot automatically capture. This indeed improves the accuracy of malware classification while reducing the complexity of requiring human experts to design features when using the classical methods (e.g., \cite{wu2023droidrl}).

Much existing ML-based malware detection research follows a fragmented approach in which researchers select specific dataset configurations, e.g., using public \cite{grosse2016,agarap2019} or private datasets \cite{dynamicdb,7552276} using static \cite{9800057,ali2020malgra,Andrioddb} or dynamic \cite{jeon2021hybrid,9092568} analysis methods, or using the pre-analysis features from existing papers \cite{12857713}. The researchers then develop models claiming superior accuracy within narrow constraints, i.e., their chosen dataset type, analysis methodology, and evaluation metrics, without demonstrating generalizability across different operational environments. This has produced numerous isolated solutions while leaving significant research gaps in systematic guidance for optimal selection of ML-techniques across diverse contexts and constraints.


We argue that building an effective machine learning-based malware detection system requires navigating a complex landscape of technical choices, including feature extraction, feature selection, and deep learning hyperparameter tuning. In practice, these variables are highly interdependent: the choice of deployment platform constrains feasible model architectures, which in turn dictate the appropriate data preprocessing pipelines and viable training strategies. Without a structured methodology, practitioners must manage these cascading decisions in an ad-hoc manner, often resulting in suboptimal configurations that fail to balance detection performance, resource consumption, and long-term maintainability. This systemic lack of guidance raises a fundamental research question: \emph{what constitutes the most suitable machine learning methodology for malware detection given varying organizational contexts, data characteristics, and operational constraints?}

To answer this, we propose a Framework for Decision-making to build ML-based Malware detection systems (FDM). The framework provides systematic guidance for malware prevention teams in constructing detection systems suited to diverse organizational constraints and requirements, offering a principled approach to methodology selection rather than another isolated solution.

The main contributions of this paper are as follows:
\begin{enumerate}
    \item A quantifiable decision-making framework (FDM) that encodes five operational input dimensions (Q1--Q5) into a Weighted Configuration Compatibility Score (WCCS), producing ranked ML configuration recommendations across nine deployment dimensions, spanning data acquisition and feature extraction to model selection and update strategy.
    \item Empirical benchmarks of classical and deep learning models across five experimental tasks: binary API-based malware classification, multi-class API-based classification, class-incremental learning, transfer learning across three heterogeneous datasets, and autoencoder-based feature extraction. Measurements cover test accuracy, AUC, training time, inference speed, peak RAM, and model size (parameter count), enabling direct comparison of model suitability under different operational constraints.
    \item A unified FDM validation table mapping each experimental finding to specific FDM input dimensions and recommendation codes, demonstrating that no universal ``one-size-fits-all" machine learning setup exists for detecting malware, which is precisely the condition the FDM is designed to address.
    \item Three illustrative deployment scenarios that demonstrate how the FDM produces internally consistent, context-appropriate configurations for IoT edge devices, ransomware defence systems, and critical-infrastructure endpoints.
\end{enumerate}

The remainder of this paper is organised as follows. Sections \ref{sec:background} and \ref{sec:relatedwork} cover the background concepts and review related work. Section \ref{sec:methodology} introduces the research methodology. Section~\ref{sec:fdm} details the proposed FDM framework, followed by the experimental evaluation and results in Section \ref{sec:experiments}. Finally, Sections \ref{sec:discussion} and \ref{sec:conclusion} discuss the findings, limitations, and the concluding remarks.

\section{Background} \label{sec:background}


In this section, we first provide background knowledge about malware analysis techniques (\ref{sec_2.3_features}). Then, we discuss malware detection approaches: signature-based, behavior-based, and heuristic-based (\ref{sec_2.2_detection_approach}). After that, we describe the malware signature creation process for traditional antivirus systems (\ref{sec_2.1_process_signature}), and the end-point protection maintenance cycle, which is an essential component of the deployment phase to ensure systems can adapt to both known and unknown threats (\ref{sec:endpoint}).





\subsection{Malware analysis techniques}\label{sec_2.3_features}


To examine files and obtain features, software analysis is carried out. The analysis is typically divided into two categories: static and dynamic analysis \cite{dynamicdb}, each with distinct advantages and specific use cases due to the complementary strengths of these approaches, details as follows:


\subsubsection{Static analysis}
When performing a static analysis, one examines the binary structure and metadata of a program without executing it to create a dataset by extracting features such as import/export tables, section headers, API calls, strings, and file metadata from PE headers, and then applies classification methods (such as rules, machine learning models, or signature matching) to these features to determine the potential maliciousness of the file. Alternatively, a software disassembling technique can be used to extract information, such as the opcode and register records, from the assembly code. An example of research that uses static analysis is Gülmez et al. \cite{9486386}, where static analysis is carried out using the Distorm3 disassembler. They extract opcode sequences, convert them into subgraphs, and apply the Random Forest (RF) classification to detect malicious code. 


\subsubsection{Dynamic analysis}
When performing dynamic analysis, the program is executed in a virtual environment such as a debugger or a sandbox, and its real-time behavior is analyzed and recorded, such as how it interacts with other programs, what it accesses, network activity, and memory allocation. The record of this can be used to create a dataset that can be used for malware detection. An example of this is a paper by Suaboot et al. \cite{dynamicdb} where they executed software and malware in a sandbox and monitored the dynamic API calls in real time. The API calls are then exported in order to be used as the dataset to train their Sub-Curve HMM method. 

\subsection{Malware detection approaches} \label{sec_2.2_detection_approach}

To determine whether a file is malicious or benign, identify malware types (e.g., ransomware, trojan, and spyware), and enable appropriate responses, different malware detection approaches are used. There are three main categories \cite{8949524}: signature-based, behavior-based, and heuristic detection. Each has distinct strengths and limitations in identifying known and unknown threats as follows.

\subsubsection{Signature-based method}
Signature-based detection identifies malware by matching patterns of programs, or signatures, against a database of known threats, making it effective for real-time detection and generally less resource-intensive compared to other approaches. However, it struggles with variants of known malware and is less effective against unknown threats and zero-day exploits. To address these limitations, YARA \cite{yaraGettingStarted} provides a more advanced solution by allowing malware researchers to create detailed descriptions of malware based on textual or binary patterns. Each YARA rule includes a set of strings and a boolean expression, offering flexible and enhanced pattern matching that can identify complex variants not captured by traditional signature-based methods.

\subsubsection{Behavior-based}
Behavior-based detection takes a different approach by focusing on the real-time actions of software, allowing it to identify malware based on its behavior rather than relying on known signatures. This method can detect new or previously unseen threats by observing malicious activities, regardless of whether the specific malware variant has been encountered before. Although behavior-based is the most flexible when considering unseen malware, compared to signature and heuristic-based methods, the behavior-based detection can be more resource-intensive and may result in a higher rate of false positives, as legitimate software might exhibit suspicious behaviors.
 
\subsubsection{Heuristic-based}
In general, heuristic-based detection serves as a more generalized approach compared to signature-based detection, enabling the identification of variants of known malware, even if these specific versions have not been previously encountered. Heuristic detection uses a plethora of algorithms \cite{yunmar2024hybrid} and rules \cite{mehtab2020addroid} to evaluate the characteristics and behaviors of applications in order to screen for suspicious activity or sections of an application's data that would point to malicious intent. 

\begin{figure}[ht]
	\centering
	\includegraphics[width=\columnwidth]{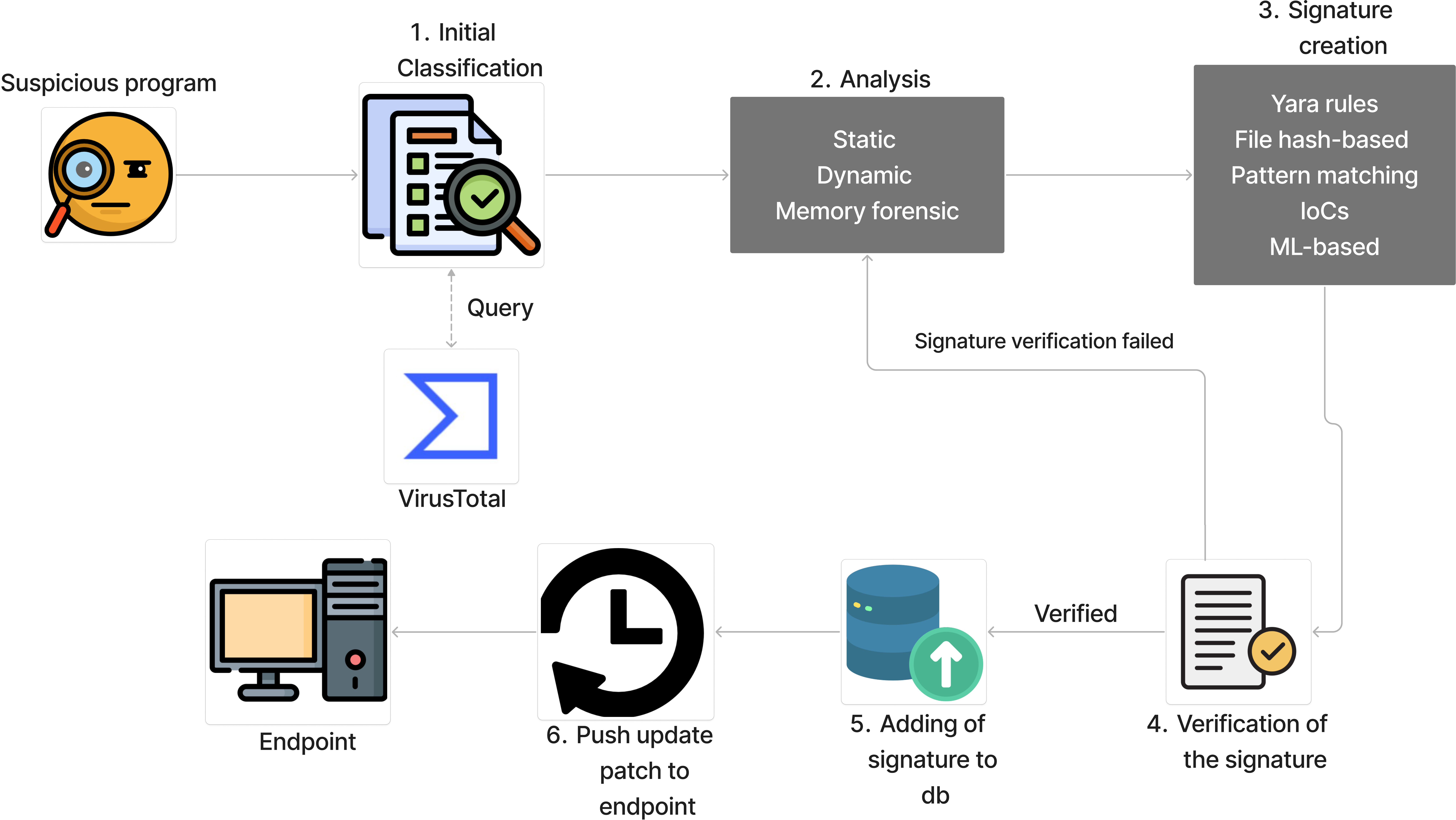}
	\caption{Processes of creating a signature for traditional signature-based antivirus systems.}
	\label{fig:EDR}
\end{figure}

\subsection{Process of malware signature creation for traditional antivirus systems} \label{sec_2.1_process_signature}
When creating malware signatures for updating the endpoint protection, the steps involved are initial classification, analysis, signature creation, signature verification, adding of signature to the database, and pushing an update patch to the endpoint, which is shown in Fig.~\ref{fig:EDR}.

\begin{enumerate}[label=Step~\arabic*:, leftmargin=*, align=left]
    \item \textit{Initial classification}: For a program under review or labeled as suspicious, the program's hash (SHA-1, SHA-256, or MD5) is first checked on platforms such as VirusTotal to determine whether a report already exists.

    \item \textit{Analysis}: If a report on the application can not be found on platforms like VirusTotal, then different types of analysis, including Static, Dynamic, and Memory forensics, are conducted on the application. From this analysis, the researcher gathers information on the suspicious activity of the application using tools like a disassembler and a sandbox.

    \item \textit{Signature creation}: The information that was gathered is then used to create a signature using differing approaches such as Yara rules, File hash-based, Pattern matching, Indicators of Compromise (IoCs), and Threat Intelligence Feeds.

    \item \textit{Signature verification}: The confirmation of the signature's accuracy in representing the program and that it does not flag the wrong program. If the signature does not accurately represent the program, then in this process, the researcher has to redo the analysis of the program (Step 2).

    \item \textit{Adding of signature to database}: Once the signature has been verified then it is added to the database of the systems alongside other signatures.

    \item \textit{Push update patch to endpoint}: The signature is then updated on the endpoint device via a patch that is released by the administrator, including future updated signatures from when new variants have emerged.


\end{enumerate}

\subsection{Endpoint protection} \label{sec:endpoint}

 \begin{figure*}[ht]
	\centering
	\includegraphics[width=0.8\textwidth]{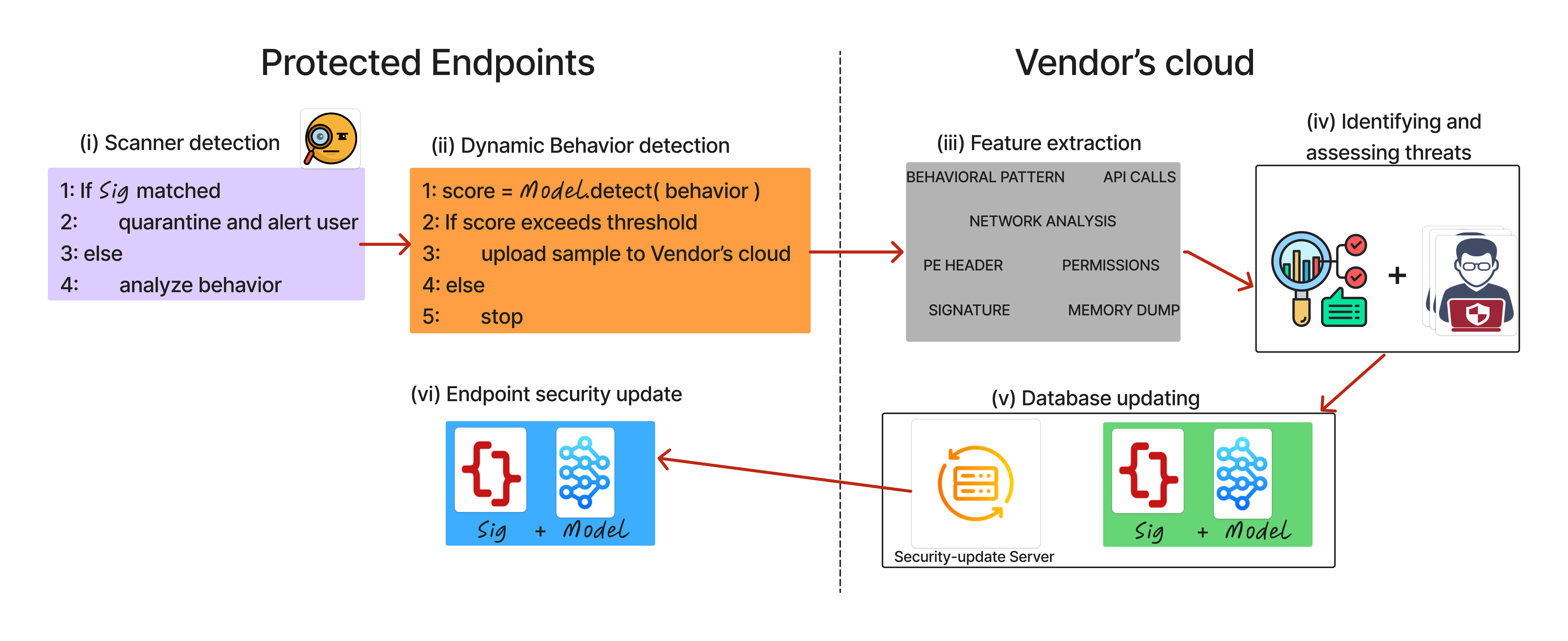}
	\caption{Modern endpoint protection: automated ML-based deployment and maintenance pipeline.}
	\label{fig:deploy}
\end{figure*}

While the previous section detailed the traditional, reactive process of creating individual malware signatures, modern endpoint protection must go further to address rapidly evolving unseen threats. Through an automated maintenance pipeline, modern endpoint protection combines endpoint-based scanning with advanced cloud-based analysis to continuously push security updates to endpoints, ensuring defense against newly detected malware.

Fig.~\ref{fig:deploy} illustrates the six-stage deployment and maintenance pipeline. On the protected endpoint, a \textit{scanner} (i) first checks incoming files against known signatures; on a match, the file is quarantined and the user is alerted. If no signature matches, \textit{dynamic behavior detection} (ii) scores the sample using a local ML model and, when the score exceeds a threshold, uploads the sample to the vendor's cloud for deeper analysis. In the cloud, \textit{feature extraction} (iii) derives static and dynamic feature sets (API calls, PE header fields, memory dumps, network traffic, permissions, behavioral patterns, and signatures) which feeds into \textit{threat identification and assessment} (iv). Here, automated analysis and reverse engineering tools help analysts identify new threats. After that, the confirmed threats trigger \textit{database updating} (v), where new signatures and an updated model are committed to the security-update server. Finally, the cycle closes with an \textit{endpoint security update} (vi) that pushes the updated signature database and model back to all protected endpoints.

\section{Related Work}\label{sec:relatedwork}

This section reviews the literature across three areas relevant to the FDM: (a)~ML detection approaches that collectively motivate the need for systematic configuration guidance; (b)~multi-criteria decision-making (MCDM) methods; and (c)~context-aware deployment frameworks that inform the FDM's design.

\subsection{ML Approaches for Malware Detection}

Numerous papers propose specific ML configurations evaluated on narrow benchmark conditions. Bayazit et al.~\cite{9800057} evaluate recurrent deep learning models on a static-feature Android dataset~\cite{8585560} (396 malware, 1,126 benign samples), finding BiLSTM outperforms LSTM, GRU, and RNN. Notably, the paper dismisses classical ML without empirical comparison, illustrating the fragmented nature of the literature. Kalash et al.\ \cite{8328749} demonstrate that CNN-based image classification achieves 98.52\% accuracy on Malimg and 99.97\% on the Microsoft dataset, establishing image-based malware visualisation as a viable detection feature. Zhangjie et al.\ \cite{jcs.2021.016632} augment LSTM training with GAN-generated malware samples, reaching 99.94\% on augmented data but 86.5\% on novel real-world samples; this gap highlights overfitting risk when the training distribution does not match deployment. Wong et al.\ \cite{9631209} combine ShuffleNet and DenseNet-201 for feature extraction with SVM classification across four image datasets, achieving 85.79–99.14\% accuracy depending on dataset complexity. Sudhakar et al.\ \cite{SUDHAKAR2021334} fine-tune ResNet50's final layer on Malimg (99.18\% accuracy, 5.14\,ms inference), demonstrating that lightweight transfer learning can match from-scratch training. Owoh et al.\ \cite{fi16100369} propose a hybrid GRU–GAN model for dynamic API call sequences, achieving 98.2\% accuracy with reduced memory overhead compared to standalone deep models.

These studies collectively demonstrate that high detection accuracy is achievable across diverse configurations. However, each paper optimises within a single combination of feature type, dataset, and model architecture, leaving practitioners without guidance on which configuration to select when their deployment context differs (in platform, resources, or update requirements) from the experimental setting.

\subsection{Multi-Criteria Decision Making for ML Configuration Selection}
\label{subsec:mcdm}

The challenge of selecting among competing ML configurations under multiple, sometimes conflicting criteria has been addressed in various domains using MCDM methods. Research by Kumar and Kaur~\cite{kumar2024mcdm} applied a fusion of WSM, TOPSIS, and VIKOR to rank algorithms for medical prediction, showing that formalizing evaluation criteria produces insightful, reproducible recommendations. In the context of modern deployment, configuration selection must increasingly account for ``platform-aware'' and ``hardware-aware'' constraints \cite{dong2024automated}. Recent taxonomies in Neural Architecture Search have shifted focus toward multi-objective optimization that treats latency, memory constraints, energy consumption, and accuracy as primary objectives \cite{benmeziane2021hardware, marculescu2018hardware}. This is essential for ``Green Machine Learning,'' where the environmental cost of a model is weighed against its classification effectiveness to ensure efficient deployment \cite{santos2026systematic}. As Menghani \cite{menghani2023efficient} highlights, practitioners must identify ``Pareto-optimal'' models that provide the highest possible accuracy under specific resource constraints, such as a maximum RAM limit or execution time. On the other hand, cross-level optimization surveys \cite{liu2023enabling} highlight that achieving resource efficiency in IoT systems requires selecting configurations that align with the limitations of the deployment hardware. Furthermore, the application of multi-criteria decision-making frameworks \cite{you2015decision} in adjacent domains, such as precision marketing, has demonstrated the effectiveness of translating qualitative operational requirements into quantifiable evaluation scales to systematically guide complex selection processes.

The FDM differs from these generic MCDM applications by providing a specialized framework for malware detection, where input dimensions (Q1--Q5) and recommendation codes (P1--P5) are grounded in empirical experiments. Unlike general weighting schemes, the WCCS scoring enables finer-grained alignment between a practitioner's operational context, such as limited hardware or strict update requirements, and the resulting configuration recommendations.

\subsection{Context-Aware and Adaptive ML Deployment}
\label{subsec:contextaware}

Existing work highlights the requirement of adapting deployed malware detection systems to varying operational contexts. For example, systems like CASANDRA \cite{narayanan2017context} employ online learning to address population drift, while others \cite{feng2020performance} focus on strict hardware constraints, such as optimizing models for varying RAM capacities or deploying ``TinyML'' configurations for IoT environments \cite{alwaisi2024securing}. Recent frameworks extend this to dynamic runtime adaptation, balancing accuracy with end-to-end latency \cite{merluzzi2021wireless, xu2023edge} or switching architectures to maintain energy efficiency (e.g., EcoMLS \cite{tedla2024ecomls}). 
%
%
While adaptive systems effectively manage runtime conditions, they assume that the foundational model architecture, feature type, and training strategy have already been established. 

The FDM is complementary to these adaptive systems: it targets the configuration selection phase that precedes deployment. By quantitatively mapping operational contexts to specific pre-deployment strategies, the FDM provides the principled baseline choices required by systems like CASANDRA and EcoMLS, addressing the configuration gap they leave open.

\section{Research Methodology}\label{sec:methodology}

This study follows a two-phase design. In the first phase, the FDM is conceptually developed: operational input dimensions (Q1--Q5) are identified from a systematic analysis of the factors that practitioners report as most influential in malware detection system design, and the WCCS scoring mechanism is formalised following established multi-criteria decision analysis principles~\cite{you2015decision}. In the second phase, four empirical experiments on three labelled datasets are conducted to validate the FDM's recommendations and provide the evidence base for its configuration codes.

The study is guided by five research questions:
\begin{itemize}
  \item \textbf{RQ1:} How do classical (RF, XGBoost, SVM) and deep learning (LSTM, BiLSTM) models compare in binary API-based malware classification when long sequences are standardised by splitting into fixed-length segments? 
  \item \textbf{RQ2:} Does the relative ranking of classical and deep models change in the more demanding multi-class setting, and what explains any reversal?
  \item \textbf{RQ3:} What are the training-time and accuracy costs and benefits of transfer learning (ImageNet pre-training) versus training from scratch, and do these vary with input modality? 
  \item \textbf{RQ4:} How effectively does class-incremental learning allow models to integrate new malware families without full retraining, and which model type best resists catastrophic forgetting? 
  \item \textbf{RQ5:} Can autoencoder-based feature extraction reduce inference latency and model complexity with minimal accuracy cost, and under what operational constraints is this trade-off justified? 
\end{itemize}

The experimental answers to RQ1--RQ5 are synthesised in Section~\ref{subsec:findings} to produce a validated FDM configuration mapping, demonstrating that the FDM's WCCS scores correctly predict which configurations dominate under each set of operational constraints.

\section{Proposed Framework for Decision-making to build ML-based Malware detection systems (FDM)}\label{sec:fdm}


This section proposes the FDM, which incorporates a quantifiable recommendation algorithm that maps the practitioner's operational goals (expressed as numeric input parameters) to a ranked set of technical configuration recommendations. The proposed algorithm applies a Weighted Configuration Compatibility Score (WCCS), a mathematically grounded multi-criteria scoring method~\cite{you2015decision}, to measure how well each candidate configuration matches the stated operational requirements.

\subsection{System Development Workflow}

\begin{figure}[ht]
    \centering
    \includegraphics[width=\columnwidth]{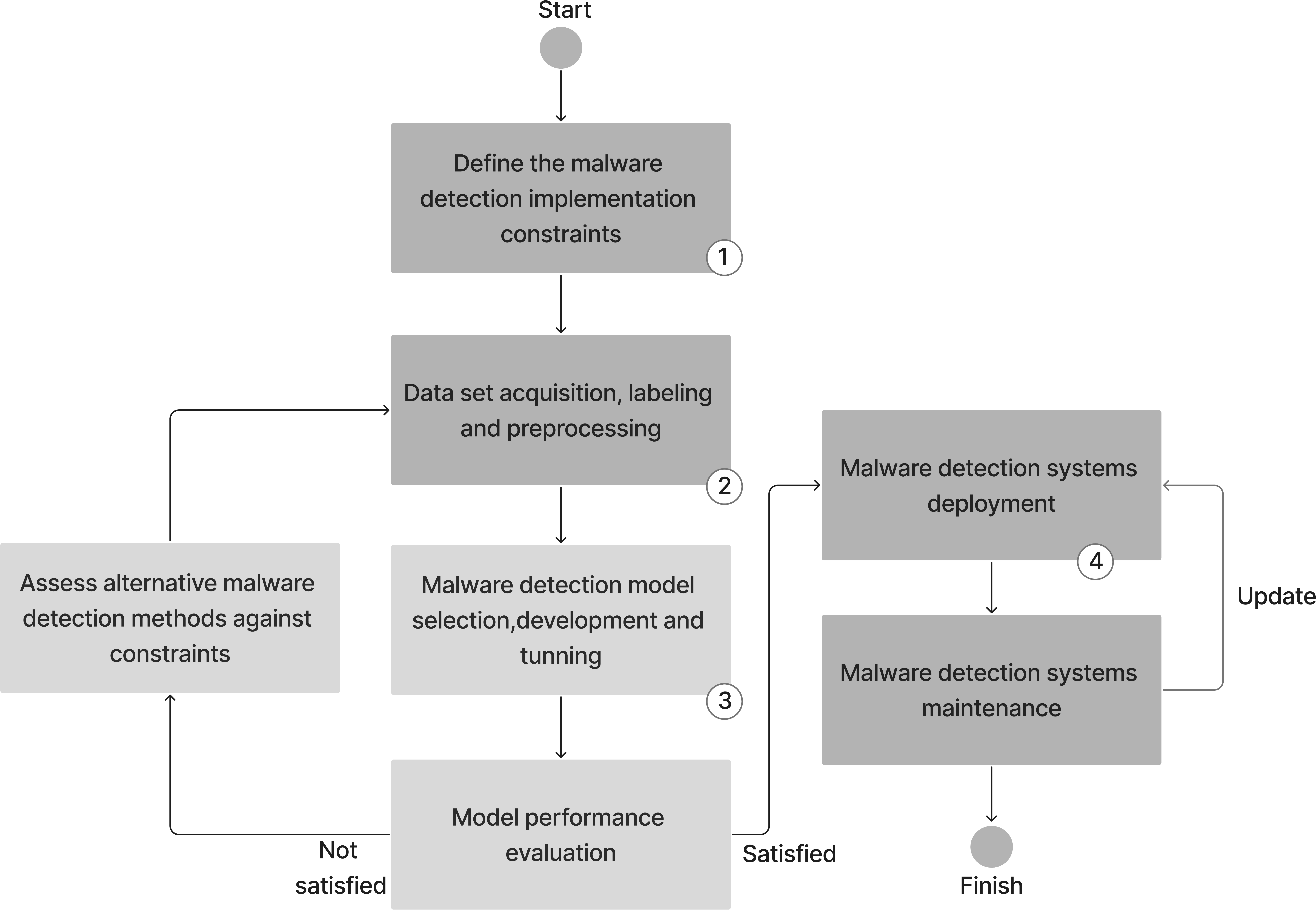}
    \caption{ML-Based Malware Detection Development Lifecycle.}
    \label{fig:DMP}
\end{figure}

In general, to develop an ML-based method for malware detection, researchers need to make a series of decisions that impact the overall efficiency and effectiveness of the method. Fig.~\ref{fig:DMP} shows the basic decision-making process of developing ML-based malware detection systems.

\subsubsection{Requirement gathering and analysis}
Similar to software development, when designing a malware detection system, it is crucial to gather requirements, constraints and define goals the organization wants to achieve. This process is crucial as it affects many of the decisions in the rest of the steps. This step is shown in Fig.~\ref{fig:DMP}, \circled{1}.

\subsubsection{Dataset acquisition}
In Fig.~\ref{fig:DMP}, \circled{2}, after requirement gathering and analysis, the next process is data acquisition, which encompasses feature extraction and engineering, feature scaling and selection, and data processing and shaping to obtain a usable dataset. The process involves: (1)~\textit{Sample collection}: gathering malware samples and benign files from sources such as VirusTotal and malware repositories; (2)~\textit{Pre-processing}: removing irrelevant features, eliminating redundancy, and normalizing data to ensure consistency; (3)~\textit{Feature extraction}: choosing the analysis approach (static, dynamic, or hybrid) based on classification goals; (4)~\textit{Feature representation}: converting extracted features into graph-based, vectorized, numerical, or embedded representations; and (5)~\textit{Feature selection}: using statistical and ML techniques such as correlation analysis, PCA, and Chi-square tests to retain the most discriminative features.

\subsubsection{Machine learning model development and evaluation}
Fig.~\ref{fig:DMP}, \circled{3}. Setting up ML models is critical for malware classification because the decisions made during development substantially affect classification performance. The process comprises: data loading and labeling; pre-processing raw data for efficient model ingestion; dataset splitting for training, validation, and testing; algorithm selection aligned with research objectives and dataset characteristics; model training; hyperparameter optimization using automated tools such as Optuna~\cite{akiba2019optuna}; and model evaluation using accuracy, precision, recall, F1-score, and ROC-AUC. However, if the model's performance is not satisfactory, the developer could try alternative ML models until the desired evaluation criteria are met.

\subsubsection{Deployment and maintenance}
Once the developer has successfully built the ML model for their specific need, it is deployed in a real environment. Here, deployment strategies differ depending on the target platform, computational resources, and real-time detection requirements. Three primary strategies exist: \textit{on-device deployment}, where the detection method runs locally on endpoint hardware (e.g., Windows, Android) for real-time classification; \textit{cloud-based deployment}, where samples are uploaded to cloud servers when on-device resources are insufficient; and \textit{hybrid deployment}, combining both. Ongoing maintenance involves periodic model retraining on updated datasets. Depending on the organization's requirements, various malware signature update strategies will be determined to make sure the system is resilient against emerging evasion techniques.

\subsection{FDM Components}
\label{subsec:fdm_components}

\begin{figure}[t]
    \centering
    \includegraphics[width=0.3\textwidth]{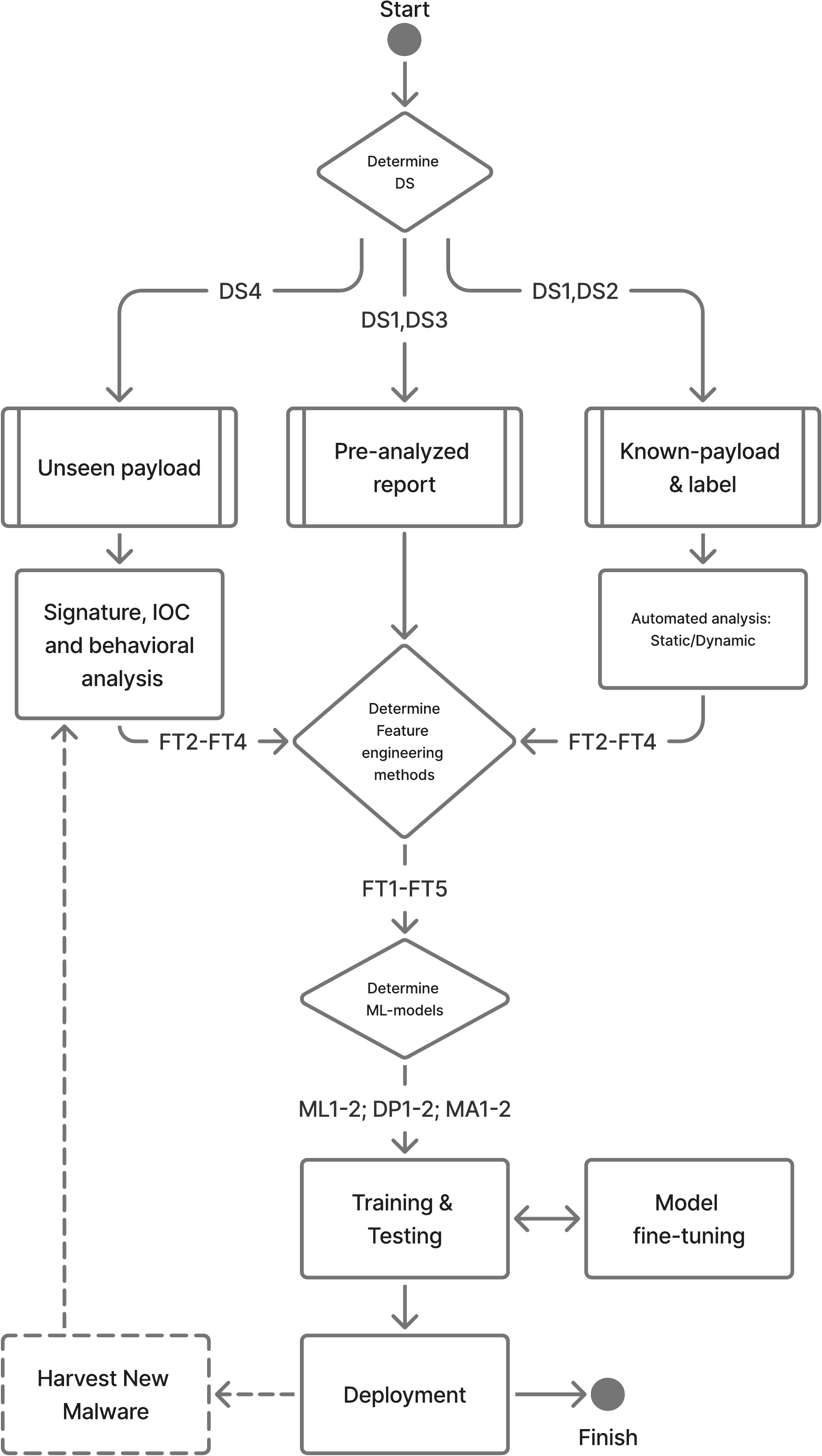}
    \caption{Decision-making flow of the FDM, showing the ML-based malware detection system development process. Each node represents a configuration option: DS=Dataset Source, FT=Feature Type, ML=ML Model, DP=Data Pre-processing approach, MA=Model Maintenance technique.}
    \label{fig:FDM_diagram}
\end{figure}

This section describes the components of the proposed FDM. Fig.~\ref{fig:FDM_diagram} illustrates the decision flow connecting all components through the FDM algorithms to determine the following five key configuration categories (DS, FT, ML, DP, and MA).

\subsubsection{Dataset Sources (DS)}
\begin{itemize}
    \item \textit{DS1 -- Secondary data, pre-reports and labels:} Sources providing reports of seen payloads, which consist of labels and a behavioral report. This might include raw behaviors such as list of API calls, mutex objects, APK permissions and so on, which are secondary information of the malware. VirusTotal \cite{yaramahdi2024detection} is the primary example; it is used for labeling and behavioral pre-analysis. Dadkhah et al. \cite{dadkhah2026cic} provide multiple layers of malware behaviors including network traffic, system calls, and CPU utilization. This type of data source is safe and could be effective to develop efficient malware detection models as it combines multiple views of malware.
    
    \item \textit{DS2 -- Actual payloads and labels:} Sources providing a wide variety of pre-labeled file types (e.g., \texttt{.exe}, \texttt{.pdf}, \texttt{.xlsm}) and scripts (e.g., \texttt{.vbs}, \texttt{.js}, \texttt{.ps1}). VirusShare \cite{virusshare} is the primary example, providing a private repository of pre-labeled malware samples. Although this data source is the most dangerous (sophisticated malware could escape from an improperly configured environment), it is the most effective way for a malware analyst to investigate payload behaviors in depth and discover more comprehensive characteristics.
    
    \item \textit{DS3 -- Labeled features:} This type is the most convenient data source for an AI engineer rather than a specialized malware analyst. It provides features extracted from processed samples (e.g., BODMAS \cite{yang2021bodmas}, EMBER \cite{anderson2018ember}, and other datasets publicly available in Kaggle and Github), enabling researchers to perform additional feature engineering.
    
    \item \textit{DS4 -- Unseen payloads:} An in-house harvesting of unknown payloads with indicators of compromise (IoC) collected from endpoint devices. These samples require dataset preparation before integration. This data source requires a malware analyst to be involved in building a signature, IoC indicators, and a behavioral report.
\end{itemize}

\subsubsection{Feature Types (FT)}
\begin{itemize}
    \item \textit{FT1 -- Image-based features:} Binary files converted into images, enabling the use of transfer learning from pretrained models, specifically, those that are designed to extract and process visual features, such as CNN and autoencoders.
    
    \item \textit{FT2 -- Sequence features
    	:} Behavioral sequences extracted via dynamic analysis, capturing API call patterns and temporal correlations.
    
    \item \textit{FT3 -- Graph features:} A Control Flow Graph (CFG) is a graph-based representation of function-call relationships, suited to graph neural network (GNN) classifiers. CFGs are usually extracted from static analysis, e.g., by decompiling the executable into high-level code from which the CFG is generated. 
    
    \item \textit{FT4 -- Static features:} Various features such as PE headers, byte histograms, and printable strings extracted directly from binaries without execution, which can be obtained from static analysis procedures.
    
    \item \textit{FT5 -- Latent features:} When using an autoencoder (e.g. \cite{zhou2017anomaly}) to process malware data (like API call sequences, byte frequencies, or image-converted binaries), the features extracted from the middle layer are latent features or encoded features. The latent feature is the compressed, hidden variables that capture the underlying structure of the malware data.
    
\end{itemize}

\subsubsection{Machine Learning Models (ML)}
\begin{itemize}
    \item \textit{ML1 -- Deep learning:} CNN architectures (EfficientNetB0, MobileNetV2) and recurrent models (LSTM, BiLSTM) for image and sequence classification, respectively.
    \item \textit{ML2 -- Classical and ensemble ML:} RF, XGBoost, SVM and voting classifiers; preferred when computational resources are limited.
\end{itemize}

\subsubsection{Data Pre-processing Approaches (DP)}
\begin{itemize}
    \item \textit{DP1 -- Sequence standardization:} 
    Partitioning variable-length data streams or sequences into fixed-length segments. This establishes a uniform input format required by various sequence-based deep and non-deep ML techniques.
        
    \item \textit{DP2 -- Dimensionality reduction:} 
    
    Utilizing feature compression techniques (e.g., autoencoder, PCA and t-SNE) to map high-dimensional inputs into a compact latent representation. This significantly reduces downstream computational complexity and test-time inference latency, making the pipeline highly suitable for deployments with strict resource constraints.

\end{itemize}

\subsubsection{Model Maintenance Techniques (MA)}
\begin{itemize}
	 \item \textit{MA1 -- Transfer learning:} Adapts model weights pretrained on large-scale external datasets (e.g., ImageNet) to new malware data distributions. This approach serves as a rapid model initialisation strategy when labelled malware data is scarce, and acts as an efficient periodic updating mechanism to significantly reduce retraining time and computational resource consumption.
	 	 
    \item \textit{MA2 -- Class-incremental learning:} Facilitates the continuous addition of new malware families to an existing model without requiring full retraining. This strategy governs both the initial model training for extensibility and the subsequent updating mechanism, effectively mitigating catastrophic forgetting while retaining prior knowledge of older threats.
    
   
\end{itemize}

\subsection{Input and Output}
\label{subsec:input_params}

The FDM defines system inputs (i.e., system's requirements) as five quantifiable parameters (i.e., Q1--Q5), and nine recommendations for configurable aspects of the malware detection system, including development and deployment, i.e., as framework's outputs. Each requirement of the system maps a practitioner's operational goal to a numeric or ordinal scale, enabling the WCCS algorithm to compute a mathematically comparable score across all configuration candidates. Tables~\ref{tab1:input_params} and \ref{tab2:output} define the input and output of the FDM, details as follows.

\begin{table*}[ht]
\centering
\caption{Framework's Inputs: Definition and Scale of the Quantifiable Parameters}
\label{tab1:input_params}
\begin{tabular}{|p{0.7cm}|p{2.8cm}|p{6.5cm}|p{4.5cm}|}
\hline
\textbf{Param.} & \textbf{Name} & \textbf{Description} & \textbf{Scale / Levels} \\ \hline
Q1 & Platform Constraint Index (PCI) & Capability and resource limits of the target deployment platform. & 1 = Mobile/Edge device (e.g., Android gateway)\newline 2 = Server or local cloud\newline 3 = Unrestricted high-performance cluster \\ \hline
Q2 & Resource Budget Index (RBI) & Available memory, compute, and GPU resources for training and inference. & 1 = Constrained ($<$4\,GB RAM, CPU-only)\newline 2 = Moderate (4--16\,GB RAM, optional GPU)\newline 3 = High-performance ($>$16\,GB RAM, dedicated GPU) \\ \hline
Q3 & Response Latency Level (RLL) & Required inference speed for the deployment use-case. & 1 = Offline/batch (hours acceptable)\newline 2 = Near-real-time (seconds acceptable)\newline 3 = Strict real-time ($<$100\,ms required) \\ \hline
Q4 & Update Frequency Score (UFS) & Required cadence for retraining or updating the model and dataset. & 1 = One-time (interval $>$3 months)\newline 2 = Periodic (weekly to monthly)\newline 3 = Continuous (daily or event-triggered) \\ \hline
Q5 & Sensitivity Ratio (SR) & Relative operational cost of a false negative vs.\ a false positive; SR = Cost(FN)/Cost(FP). & Low: SR $<$ 1.0 (FP-sensitive)\newline Medium: $1.0 \le \text{SR} \le 3.0$ (balanced)\newline High: SR $>$ 3.0 (FN-sensitive) \\ \hline
\end{tabular}
\end{table*}

\begin{table*}[h]
	\centering
	\caption{Framework's Outputs: Recommendation Codes and Quantifiable Scales}
	\label{tab2:output}
	\resizebox{\textwidth}{!}{
		\begin{tabular}{|p{0.7cm}|p{2cm}|p{4.5cm}|p{2.8cm}|p{5cm}|}
			\hline
			\textbf{Code} & \textbf{Name} & \textbf{Detail} & \textbf{Options} & \textbf{Quantifiable Scale} \\ \hline
			\rowcolor[HTML]{FFFFFF}
			P1 & Data acquisition & Suggest the dataset source based on update cadence and sensitivity requirements. & Pre-reported (DS1), Raw malware with labels (DS2), Labelled features (DS3), In-house (DS4) & Acquisition Level (AL): 1=Public, 2=Private repo, 3=In-house collection. Driven by Q4 (UFS, $w$=0.45). \\ \hline
			\rowcolor[HTML]{EFEFEF}
			
			\multicolumn{5}{|c|}{ P2: ML model fine-tunings }  \\ \hline
			\rowcolor[HTML]{EFEFEF}
			
			P2.1 & ML model & Ranked list of ML models by WCCS$_{2.1}$ compatibility score. & RF, XGBoost, CNN, LSTM, BiLSTM, GNN & Resource Score (RS): 1=Low (RF, XGBoost), 2=Medium (LSTM, BiLSTM), 3=High (CNN). Output: top-$k$ ranked list. \\ \hline
			\rowcolor[HTML]{EFEFEF}
			
			P2.2 & Hyperparameter auto-tuning & Recommend HPO tool based on model complexity and resource budget. & Optuna (default), Ray Tune, Hyperopt & Search Space Index (SSI): 1=Small ($<$10 params), 2=Medium (10--30), 3=Large ($>$30). Mapped from Q2 and model RS. \\ \hline
			\rowcolor[HTML]{EFEFEF}
			
			P2.3 & Fixed hyperparameters & Specify hyperparameters to fix to ensure reproducibility. & Epoch count, sampling rate & Epoch count: integer $\in [10, 200]$; sampling rate: float $\in [0.1, 1.0]$. Values determined by Q3 and dataset size. \\ \hline
			\rowcolor[HTML]{FFFFFF}
			
			\multicolumn{5}{|c|}{ P3: Data Pre-processing }  \\ \hline
			\rowcolor[HTML]{FFFFFF}
			
			P3.1 & Feature extraction & Specify static, dynamic, or hybrid extraction based on latency and infrastructure. & Static (ET=1), Dynamic (ET=2), Hybrid (ET=3) & Extraction Type (ET) $\in \{1,2,3\}$. Constrained by Q1 and Q3: Q1=1 (Edge) or Q3=3 $\rightarrow$ ET=1; otherwise ET $\in \{2, 3\}$. \\ \hline
			\rowcolor[HTML]{FFFFFF}
			
			P3.2 & Pre-processing & Recommend pre-processing stages: cleansing, normalization, and EDA. & Cleansing, normalization, EDA & Pipeline Stage Count (PSC): 1=Minimal, 2=Standard, 3=Full pipeline. Driven by Q3 and Q2 (autoencoder pre-processing under tight latency/budget); high Q5 (SR$>$3.0) $\rightarrow$ PSC=3. \\ \hline
			\rowcolor[HTML]{FFFFFF}
			
			P3.3 & Feature selection & Specify features from the dataset most discriminative for the target malware families. & API calls, opcodes, PE headers, byte histograms, strings & Feature Dimensionality (FD): Low ($<$100), Medium (100--1,000), High ($>$1,000). Constrained by Q2 (RBI, $w$=0.40). \\ \hline
			\rowcolor[HTML]{EFEFEF}
			
			P4 & Updating priority & Determine model/dataset refresh frequency and updating mechanism. & Continuous (MA2), Periodic (MA1), One-time & Update Interval (UI) in days: UI$\le$7 (continuous), 7$<$UI$\le$30 (periodic), UI$>$90 (one-time). Driven by Q4 ($w$=0.70). \\ \hline
			\rowcolor[HTML]{FFFFFF}
			
			P5 & Detection sensitivity & Determine the acceptable FP--FN trade-off and HPO objective. & Recall-focused, Balanced, Precision-focused & Sensitivity Ratio SR=Cost(FN)/Cost(FP): SR$>$3.0 $\rightarrow$ recall objective; $1.0 \le \text{SR} \le 3.0$ $\rightarrow$ F1; SR$<$1.0 $\rightarrow$ precision. Dominant weight Q5 ($w$=0.75). \\ \hline
		\end{tabular}
	}
\end{table*}

\begin{description}[style=multiline, labelwidth=0.7cm, leftmargin=1cm, align=left]
    \item[\textit{Q1 -- Platform Constraint Index (PCI):}] Captures the capability and resource limits of the target deployment platform, scored as 1 = mobile/edge device (e.g., Android gateway), 2 = server or local cloud, and 3 = unrestricted high-performance cluster. It constrains the feasible set of ML models (P2.1) and feature extraction methods (P3.1): edge devices (PCI=1) require lightweight models (RS=1), whereas unrestricted clusters (PCI=3) can accommodate deep CNN architectures (RS=3).
    
    \item[\textit{Q2 -- Resource Budget Index (RBI):}] Represents the available memory, compute, and GPU resources for training and inference, scored as 1 = constrained ($<$4\,GB RAM, CPU-only), 2 = moderate (4--16\,GB RAM, optional GPU), and 3 = high-performance ($>$16\,GB RAM, dedicated GPU). It governs model complexity and feature dimensionality: constrained deployments (RBI=1) favour low-dimensional features and classical ML, whereas high-performance environments (RBI=3) enable deep learning and high-dimensional feature spaces.
    
    \item[\textit{Q3 -- Response Latency Level (RLL):}] Specifies the required inference speed for the deployment use-case, scored as 1 = offline/batch (hours acceptable), 2 = near-real-time (seconds acceptable), and 3 = strict real-time ($<$100\,ms required). It determines the permissible feature extraction method: strict real-time requirements (RLL=3) mandate static analysis (ET=1) without sandboxing overhead, whereas offline deployments (RLL=1) can accommodate dynamic or hybrid analysis (ET=2 or 3).
    
    \item[\textit{Q4 -- Update Frequency Score (UFS):}] Defines the required cadence for retraining or updating the model and dataset, scored as 1 = one-time (interval $>$3 months), 2 = periodic (weekly to monthly), and 3 = continuous (daily or event-triggered). It drives data acquisition (P1) and the model updating strategy (P4): continuous-update deployments (UFS=3) require in-house or private data collection and class-incremental learning (MA2), whereas one-time deployments (UFS=1) can rely on public datasets.
    
    \item[\textit{Q5 -- Sensitivity Ratio (SR):}] Quantifies the relative operational cost of a false negative versus a false positive, defined as SR = Cost(FN)/Cost(FP) and graded as low (SR $<$ 1.0, FP-sensitive), medium ($1.0 \le \text{SR} \le 3.0$, balanced), and high (SR $>$ 3.0, FN-sensitive). High-SR environments, such as critical infrastructure, prioritise recall and trigger full preprocessing pipelines, whereas low-SR environments prioritise precision.
\end{description}

The recommendation codes (P1--P5) in Table~\ref{tab2:output}, column \textit{Quantifiable Scale}, suggests the system's configurations based on the value calculated from the compatibility functions detailed in the following Section~\ref{subsec:wccs}.


\subsection{Weighted Configuration Compatibility Scoring (WCCS)}
\label{subsec:wccs}

The WCCS formalizes the mapping from operational goals to ML configuration choices. For each recommendation dimension $k$ ($k \in \{\text{P1, P2.1, P2.2, P2.3, P3.1, P3.2, P3.3, P4, P5}\}$) and each candidate configuration option $o \in \Omega_k$, the WCCS is defined as:

\begin{equation}
\mathrm{WCCS}_k(o) = \sum_{j=1}^{5} w_{kj} \cdot c_{kj}(Q_j, o)
\label{eq:wccs}
\end{equation}

\noindent where $\mathrm{WCCS}_k(o) \in [0,1]$ is the overall compatibility score that option $o$ achieves for recommendation area $k$, with higher values indicating a better-suited option; $w_{kj} \in [0,1]$ is the importance weight of input dimension $Q_j$ for recommendation area $k$ (with $\sum_j w_{kj} = 1$), and $c_{kj}(Q_j, o) \in [0,1]$ is the compatibility of option $o$ with requirement value $Q_j$. The set $\Omega_k$ comprises the candidate configuration options considered for area $k$ (for instance, the candidate ML models for P2.1), and $o$ denotes an individual option drawn from this set.

Intuitively, Eq.~\ref{eq:wccs} scores each candidate option as a weighted average of its compatibility with the five operational requirements. The compatibility term $c_{kj}(Q_j,o)\in[0,1]$ measures how well option $o$ meets the $j$-th requirement, ranging from a complete mismatch ($0$) to a perfect fit ($1$), while the weight $w_{kj}$ expresses the relative importance of that requirement for decision $k$. Because the weights are normalised to sum to one, more relevant requirements contribute proportionally more to the total, so an option that aligns with the higher-weighted requirements can outrank one that is strong on only a peripheral criterion.

The recommended configuration for area $k$ is then:

\begin{equation}
o^*_k = \operatorname*{argmax}_{o \in \Omega_k}\; \mathrm{WCCS}_k(o)
\label{eq:recommendation}
\end{equation}

\noindent where $o^*_k$ denotes the recommended option for area $k$, that is, the candidate in $\Omega_k$ that maximises the WCCS.

Eq.~\ref{eq:recommendation} then selects, for each decision $k$, the option that attains the highest score; the $\operatorname*{argmax}$ operator returns the maximising option $o \in \Omega_k$ itself rather than the score value. Performing this selection across all decision areas (P1--P5) yields the complete configuration that the FDM recommends for a given requirement vector $\mathbf{Q}$.

The compatibility function $c_{kj}(Q_j, o)$ encodes domain knowledge: for numeric dimensions Q1--Q4, it uses a piecewise linear scoring function that assigns 1.0 when an option perfectly suits the requirement level and decreases linearly toward 0.0 as the mismatch grows. For Q5 (SR), a monotonic sigmoid-like mapping is applied so that recall-oriented models score highly when SR is large, while precision-oriented configurations are preferred when SR is small. This approach is more robust than simple if-else branching because it naturally handles partial matches: a configuration that is not optimal on every dimension but scores highly overall may be more practical than one optimal on only one dimension.

Table~\ref{tab:weight_matrix} presents the importance weight matrix $w_{kj}$. The dominant weights reflect the primary driver of each recommendation: Q4 (UFS) dominates P4 ($w = 0.70$) since update frequency directly determines the updating strategy; Q5 (SR) dominates P5 ($w = 0.75$) since the sensitivity ratio directly governs the FP--FN trade-off objective; and Q1 (PCI) is the strongest driver of P2.1 ($w = 0.30$) because the deployment platform immediately constrains the feasible model set. The remaining weights are calibrated to the empirically validated input--output associations reported in Table~\ref{tab:fdm_validation}: Q1 and Q2 jointly drive model selection (P2.1); Q3 (latency) and Q1 (platform) govern feature extraction (P3.1); Q2 and Q3 govern pre-processing (P3.2); Q2 constrains feature selection (P3.3); and Q4, with Q2 as a secondary factor, drives the update strategy (P4).

\begin{table*}[ht]
\centering
\caption{WCCS Importance Weight Matrix ($w_{kj}$); each recommendation (column) sums to 1.00.}
\label{tab:weight_matrix}
\resizebox{\textwidth}{!}{
\begin{tabular}{|l|c|c|c|c|c|c|c|c|c|}
\hline
\textbf{Input} & \textbf{P1 Data Acq.} & \textbf{P2.1 ML Model} & \textbf{P2.2 HPO} & \textbf{P2.3 Fixed HP} & \textbf{P3.1 Feat.\ Ext.} & \textbf{P3.2 Pre-proc.} & \textbf{P3.3 Feat.\ Sel.} & \textbf{P4 Update} & \textbf{P5 Sensitivity} \\ \hline
Q1 (PCI) & 0.10 & 0.30 & 0.25 & 0.10 & 0.30 & 0.05 & 0.10 & 0.05 & 0.00 \\ \hline
Q2 (RBI) & 0.10 & 0.25 & 0.40 & 0.20 & 0.10 & 0.30 & 0.40 & 0.15 & 0.05 \\ \hline
Q3 (RLL) & 0.15 & 0.15 & 0.10 & 0.40 & 0.40 & 0.35 & 0.20 & 0.05 & 0.10 \\ \hline
Q4 (UFS) & 0.45 & 0.15 & 0.10 & 0.15 & 0.10 & 0.05 & 0.10 & 0.70 & 0.10 \\ \hline
Q5 (SR)  & 0.20 & 0.15 & 0.15 & 0.15 & 0.10 & 0.25 & 0.20 & 0.05 & 0.75 \\ \hline
\textbf{Total} & \textbf{1.00} & \textbf{1.00} & \textbf{1.00} & \textbf{1.00} & \textbf{1.00} & \textbf{1.00} & \textbf{1.00} & \textbf{1.00} & \textbf{1.00} \\ \hline
\end{tabular}
}
\end{table*}

\subsection{Decision-Making Algorithms}
\label{subsec:algorithm}

\begin{algorithm} 
\caption{FDM Phase~1: Data \& Feature Configuration}\label{alg:fdm1}
\begin{algorithmic}[1]

\Require $Q_1$--$Q_5$ (PCI, RBI, RLL, UFS, SR)

\Ensure P1, P3.1, P3.2, P3.3, DP, FT

\State \textbf{// Step 1: Data Acquisition (P1)}
\If{$Q_4 = 1$}
    \State P1 $\leftarrow$ DS3 (public dataset; AL\,=\,1)
\ElsIf{$Q_4 = 2$}
    \State P1 $\leftarrow$ DS3 or DS2 (AL\,=\,1--2)
\Else
    \State P1 $\leftarrow$ DS2 or DS4 (in-house; AL\,=\,2--3)
\EndIf
\If{$Q_5 > 3.0$} \Comment{high sensitivity override}
    \State P1 $\leftarrow$ DS4 (AL\,=\,3)
\EndIf
\State \textbf{// Step 2: Feature Extraction (P3.1) constrained by platforms}
\If{$Q_1 = 1$ or $Q_3 = 3$} \Comment{Edge dev. OR strict real-time}
    \State FT $\leftarrow$ FT1 or FT4;\quad P3.1: ET\,=\,1 (Static)
\ElsIf{$Q_3 = 2$}
    \State FT $\leftarrow$ FT1, FT2, or FT4;\quad P3.1: ET\,=\,1--2
\Else
    \State FT $\leftarrow$ FT2, FT3, or FT5;\quad P3.1: ET\,=\,2--3
\EndIf
\State \textbf{// Step 3: Pre-processing \& Feature Selection}
\State P3.2: PSC $\leftarrow Q_2$; \textbf{if} $Q_5 > 3.0$ \textbf{then} PSC $\leftarrow 3$
\State P3.3: FD $\leftarrow$ Low$(Q_2{=}1)$ / Med$(Q_2{=}2)$ / High$(Q_2{=}3)$
\If{$FT = FT2$}
	\State DP $\leftarrow$ DP1 (Sequence standardization)
\ElsIf{$Q_2 = 1$}
	\State DP $\leftarrow$ DP2 (Dimensionality reduction)
\EndIf
\end{algorithmic}
\end{algorithm}

\begin{algorithm} 
\caption{FDM Phase~2: Model development \& Deployment}\label{alg:fdm2}
\begin{algorithmic}[1]
\Require $Q_2$--$Q_5$; FT, DP (from Alg.~\ref{alg:fdm1})
\Ensure P2.1, P2.2, P2.3, P4, P5; full set $\mathcal{R}$

\State \textbf{// Step 4: ML Model Selection (P2.1) via WCCS}
\ForAll{$o \in \{\text{RF, XGBoost, LSTM, BiLSTM, CNN, GNN}\}$}
    \State $\mathrm{WCCS}_{2.1}(o) \leftarrow \sum_{j=1}^{5} w_{2.1,j} \cdot c_{2.1,j}(Q_j, o)$ \hfill(Eq.~\ref{eq:wccs})
\EndFor
\State P2.1 $\leftarrow$ top-$k$ models by $\mathrm{WCCS}_{2.1}$ (descending)
\State \textbf{// Step 5: Hyperparameter Optimisation (P2.2, P2.3)}
\State SSI $\leftarrow$ RS of P2.1 model \Comment{1=small, 2=med, 3=large}
\State P2.2 $\leftarrow$ Optuna (SSI);\quad P2.3: epoch $\!\in\![10,200]$, rate $\!\in\![0.1,1.0]$
\State \textbf{// Step 6: Updating Priority (P4)}
\If{$Q_4 = 3$}
    \State P4 $\leftarrow$ MA2 (Class-Incr.);\quad UI $\le$ 7\,days
\ElsIf{$Q_4 = 2$}
    \State P4 $\leftarrow$ MA1 (TL fine-tune);\quad UI\,=\,7--30\,days
\Else
    \State P4 $\leftarrow$ none;\quad UI $>$ 90\,days
\EndIf
\State \textbf{// Step 7: Detection Sensitivity (P5)}
\If{$Q_5 > 3.0$}
    \State P5 $\leftarrow$ Recall-focused;\quad HPO obj.\,=\,Recall
\ElsIf{$Q_5 \ge 1.0$}
    \State P5 $\leftarrow$ Balanced;\quad HPO obj.\,=\,F1
\Else
    \State P5 $\leftarrow$ Precision-focused;\quad HPO obj.\,=\,Precision
\EndIf
\State \Return $\mathcal{R} = \{\text{P1, P3.1--P3.3, DP, P2.1, P2.2,}\}$
\State \phantom{\Return $\mathcal{R} = \{ $} $\text{P2.3, P4, P5}\}$
\end{algorithmic}
\end{algorithm}

The FDM procedure is split into two single-column algorithms that together implement the WCCS formulation (Eq.~\ref{eq:wccs}--\ref{eq:recommendation}). Algorithm~\ref{alg:fdm1} covers \textit{data and feature configuration} (Steps~1--3): it selects the dataset source, feature extraction method, and preprocessing pipeline driven primarily by $Q_3$, $Q_4$, and $Q_5$, and outputs the selected feature type~FT for use in Algorithm~\ref{alg:fdm2}. Algorithm~\ref{alg:fdm2} covers \textit{model selection and deployment} (Steps~4--8): Step~4 applies the full WCCS ranking to select ML models; Steps~5--6 determine the training strategy and hyperparameter search space; Steps~7--8 apply direct threshold mappings to set the update mechanism and the FP--FN trade-off objective. Each step maps to a node cluster in Fig.~\ref{fig:FDM_diagram}.



\subsection{Example Deployment Scenarios}
\label{subsec:scenarios}

To demonstrate the practical utility of the WCCS algorithm, Table~\ref{tab:scenarios} presents the complete FDM recommendation set for three representative real-world deployment scenarios: (i)~securing an IoT network, (ii)~protecting import-export business endpoints from ransomware, and (iii)~safeguarding critical electric power infrastructure. Each scenario yields a distinct input vector $\mathbf{Q}$ and a correspondingly different but internally consistent configuration recommendation, illustrating how the framework adapts to diverse operational contexts.

\begin{table*}[h]
\centering
\caption{FDM Configuration Recommendations for Three Example Scenarios}
\label{tab:scenarios}
\resizebox{\textwidth}{!}{
\begin{tabular}{|p{2.5cm}|p{4.5cm}|p{4.5cm}|p{4.5cm}|}
\hline
\textcolor{black}{\textbf{Parameter}} & \textcolor{black}{\textbf{Scenario 1: IoT Network Security}} & \textcolor{black}{\textbf{Scenario 2: Endpoint Ransomware Defence (Import-Export Business)}} & \textcolor{black}{\textbf{Scenario 3: Critical Infrastructure (Electric Power Distribution)}} \\ \hline
\textbf{Deployment context} & Fleet of IoT gateways/sensors on a factory floor; malware on any device disrupts production. & Windows workstations and file servers; ransomware encrypts shipment data irreversibly. & SCADA/ICS systems controlling power distribution; a missed detection risks grid outage. \\ \hline

\textbf{Q1 (PCI)} & 1 -- Mobile/Edge (IoT gateways) & 2 -- Server/cloud & 3 -- Unrestricted on-premise \\ \hline

\textbf{Q2 (RBI)} & 1 -- Constrained ($<$4\,GB, CPU) & 2 -- Moderate (4--16\,GB, opt.\ GPU) & 3 -- High ($>$16\,GB, GPU) \\ \hline
\textbf{Q3 (RLL)} & 3 -- Real-time ($<$100\,ms) & 2 -- Near-real-time (seconds) & 3 -- Real-time ($<$100\,ms) \\ \hline

\textbf{Q4 (UFS)} & 2 -- Periodic (weekly) & 3 -- Continuous (daily) & 3 -- Continuous (daily) \\ \hline
\textbf{Q5 (SR)} & Medium (SR$\approx$2.0) & High (SR$\approx$4.5) & High (SR$\approx$8.0) \\ \hline

\textbf{P1 Data Acq.} & Public + VirusShare (AL=1--2); periodic refresh & VirusShare + in-house captures (AL=2--3); daily pipeline & In-house ICS samples + VirusShare (AL=3); classified dataset \\ \hline

\textbf{P2.1 ML Model} & RF / XGBoost (RS=1); lightweight for edge HW & LSTM (RS=2); captures ransomware API sequences & EfficientNetB0 CNN (RS=3); highest accuracy for binary image analysis \\ \hline

\textbf{P2.2 HPO} & Optuna SSI=1 (small search space) & Optuna SSI=2 (medium search space) & Optuna SSI=3 (large search space) \\ \hline

\textbf{P2.3 Fixed HP} & Epoch: 50--100; rate: 0.5 & Epoch: 100--150; rate: 0.3 & Epoch: 150--200; rate: 1.0 \\ \hline

\textbf{P3.1 Feature Ext.} & Static (ET=1); FT4 (PE headers, byte hist.) & Dynamic (ET=2); FT2 (API call sequences) & Static (ET=1); FT1 (binary image) \\ \hline

\textbf{P3.2 Pre-proc.} & Minimal (PSC=1): cleansing & Full (PSC=3): cleansing + norm.\ + EDA & Full (PSC=3): cleansing + norm.\ + EDA \\ \hline

\textbf{P3.3 Feature Sel.} & FD=Low ($<$100): PE headers, strings & FD=Medium (100--1,000): API n-grams, opcodes & FD=High ($>$1,000): binary image pixels, PE histogram \\ \hline


\textbf{P4 Update Priority (Model Maintenance)} & MA1 (TL Fine-Tuning); UI=7--30\,days 
(Periodic) & MA2 (Class-Incremental); UI$\le$7\,days (Continuous) & MA2 (Class-Incremental); UI$\le$7\,days (Continuous)\\ \hline

\textbf{P5 Sensitivity} & Balanced (F1 objective); SR$\approx$2.0 & Recall-focused (minimise FN); SR$\approx$4.5 & Recall-focused (minimise FN); SR$\approx$8.0 \\ \hline


\end{tabular}
}
\end{table*}

The three scenarios demonstrate that the WCCS algorithm produces coherent, contextually appropriate configurations. In Scenario~1, the tight resource and latency constraints (Q1=1, Q2=1, Q3=3) drive the framework toward lightweight, static-analysis-based classifiers (RF/XGBoost on PE features) with a minimal preprocessing pipeline. In Scenario~2, the combination of high sensitivity (SR$\approx$4.5) and continuous update requirements (Q4=3) selects LSTM with API sequences, a full preprocessing pipeline, daily class-incremental updates, and a recall-maximizing HPO objective. In Scenario~3, both constraints further intensify (SR$\approx$8.0, Q1=3), enabling the most resource-intensive configuration: CNN with image-based features, full pipeline, and continuous CIL updates, all optimized for maximum recall to minimize the risk of catastrophic missed detections.

\section{Experiments and Framework Validation}\label{sec:experiments}

	All experiments were conducted on a custom desktop workstation (13th~Gen Intel Core i5-13500, 32\,GB RAM, GPU NVIDIA GeForce RTX 4060 with VRAM 8GB GDDR6) running Ubuntu~22.04.2\,LTS. Pre-processing pipelines were implemented in MATLAB~R2022b; model training and evaluation used Python~3 with Pandas, Scikit-learn, and Keras. Three labelled datasets were employed throughout:
	\begin{enumerate}
		\item \textbf{Dynamic API call dataset} \cite{dynamicdb}: a private collection of Windows API call sequences captured by dynamic execution, covering eight malware families (Hivecoin, Ramnit, FakeAV, Lokibot, Ransom, Rootkit, Keylogger, Zeus) and a benign class. Sequences exceeding 1,000 API calls are segmented into fixed-length subsequences of exactly 1,000 calls.
		\item \textbf{Malimg dataset} \cite{malimg}: a public benchmark of 9,339 malware binary files rendered as $32{\times}32$ grayscale images, drawn from 25 distinct families.
		\item \textbf{Android static API dataset} \cite{Andrioddb}: a public Android dataset containing 15,036 applications (5,560 malware, 9,476 benign), represented by permission vectors extracted from each APK's manifest file.
	\end{enumerate}
	All classical machine-learning models were hyperparameter-optimised via Optuna \cite{akiba2019optuna} (30–100 trials depending on the experiment). Deep learning models were trained with early stopping (patience = 5 epochs). Hardware and software configurations were held constant across all experiments to ensure reproducible comparisons.

\subsection{Experiment~1: API-Based Malware Classification with Sequence Splitting}
\label{subsec:exp1}

%
	This experiment evaluates \emph{sequence splitting} as a pre-processing strategy for Windows API call–based malware classification, benchmarking both classical machine learning (ML) classifiers and deep learning sequence models under identical conditions. Splitting long traces into fixed-length subsequences of 1,000 calls is consistent with prior work showing that segmented sequences yield superior learning signals compared to truncated or zero-padded alternatives \cite{vhito2023effects}. Two classification tasks are examined: (i)~binary classification (malware family vs.\ benign Windows processes) and (ii)~multi-class classification (all families simultaneously).
 
\subsubsection{Binary Classification}
Each of the eight malware families was evaluated independently against benign Windows processes. Three classical models, namely Random Forest~(RF), XGBoost, and SVM, were compared against two recurrent models, LSTM and Bidirectional LSTM~(BiLSTM).

Tables~\ref{tab:exp1_binary_classic} and~\ref{tab:exp1_binary_dl} summarise per-family and aggregated performance. Key metrics are test accuracy, AUC, and peak RAM usage.

\begin{table*}[htbp]
	\centering
	\caption{Binary malware classification: classical ML models (RF, XGBoost, SVM).
		Best value among RF, XGBoost, and SVM per family (and overall mean) in \textbf{bold} for validation, test accuracy, and AUC.}
	\label{tab:exp1_binary_classic}
	\begin{adjustbox}{width=\textwidth}
		\begin{tabular}{ll
				S[table-format=2.2]
				S[table-format=2.2]
				S[table-format=3.1]
				S[table-format=1.2]
				S[table-format=3.2]}
			\toprule
			{Model} & {Malware Family} &
			{Val Acc (\%)} & {Test Acc (\%)} & {AUC (\%)} &
			{Train Time (min)} & {RAM (MB)} \\
			\midrule
			\multirow{8}{*}{Random Forest}
			& FakeAV    & 96.79 & 95.91 & 99.55 & 1.82 & 16.56 \\
			& Hivecoin    & 98.87 & 98.26 & 99.91 & 1.18 & 0.00  \\
			& Keylogger     & \textbf{98.46} & 97.64 & 99.64 & 3.34 & 0.00  \\
			& Lokibot    & 97.78 & \textbf{97.03} & 99.65 & 1.32 & 0.00  \\
			& Ramnit  & 98.71 & 97.81 & 99.42 & 1.86 & 0.00  \\
			& Ransom & 97.31 & 96.61 & 98.69 & 5.28 & 0.00  \\
			& Rootkit    & 97.56 & 96.30 & 99.48 & 2.09 & 0.00  \\
			& Zeus    & 97.43 & 96.91 & \textbf{99.80} & 3.18 & 0.00  \\
			\cmidrule{2-7}
			& \textit{Mean} & 97.86 & 97.06 & 99.52 & 2.51 & 2.07 \\
			\midrule
			\multirow{8}{*}{XGBoost}
			& FakeAV    & \textbf{97.27} & \textbf{96.69} & \textbf{99.66} & 5.73 & 68.43  \\
			& Hivecoin    & \textbf{99.18} & \textbf{98.56} & \textbf{99.94} & 2.63 & 5.31   \\
			& Keylogger     & 98.32 & \textbf{97.84} & \textbf{99.69} & 7.38 & 28.91  \\
			& Lokibot    & \textbf{97.86} & 96.95 & \textbf{99.69} & 3.64 & 0.91   \\
			& Ramnit  & \textbf{99.10} & \textbf{98.71} & \textbf{99.87} & 1.45 & 0.47   \\
			& Ransom & \textbf{97.66} & \textbf{96.96} & \textbf{99.04} & 9.06 & 23.02  \\
			& Rootkit    & \textbf{97.82} & \textbf{96.64} & \textbf{99.61} & 6.88 & 0.31   \\
			& Zeus    & \textbf{97.75} & \textbf{97.36} & \textbf{99.80} & 12.38 & 1.70  \\
			\cmidrule{2-7}
			& \textit{Mean} & \textbf{98.12} & \textbf{97.46} & \textbf{99.66} & 6.14 & 16.13 \\
			\midrule
			\multirow{8}{*}{SVM}
			& FakeAV    & 95.03 & 94.84 & 95.14 & 2.09 & 0.31 \\
			& Hivecoin    & 96.00 & 94.56 & 95.89 & 1.84 & 0.16 \\
			& Keylogger     & 96.97 & 95.96 & 95.45 & 6.38 & 0.00 \\
			& Lokibot    & 95.80 & 94.65 & 94.97 & 2.40 & 0.00 \\
			& Ramnit  & 94.47 & 92.42 & 87.76 & 0.85 & 0.00 \\
			& Ransom & 96.61 & 96.38 & 95.40 & 9.44 & 0.00 \\
			& Rootkit    & 95.55 & 94.45 & 94.41 & 2.12 & 0.00 \\
			& Zeus    & 96.33 & 95.76 & 95.45 & 5.81 & 0.00 \\
			\cmidrule{2-7}
			& \textit{Mean} & 95.85 & 94.88 & 94.31 & 3.87 & 0.06 \\
			\bottomrule
		\end{tabular}
	\end{adjustbox}
\end{table*}

\begin{table*}[htbp]
	\centering
	\caption{Binary malware classification: LSTM and BiLSTM deep learning models.}
	\label{tab:exp1_binary_dl}
	\begin{adjustbox}{width=0.9\textwidth}
		\begin{tabular}{ll
				S[table-format=2.1] S[table-format=1.2]
				S[table-format=2.1] S[table-format=2.1]
				S[table-format=2.1] S[table-format=4.0]}
			\toprule
			{Model} & {Malware Family} &
			{Train Time (min)} & {Test Time (s)} &
			{Val Acc (\%)} & {Test Acc (\%)} & {AUC (\%)} & {RAM (MB)} \\
			\midrule
			\multirow{9}{*}{LSTM}
			& Ramnit  & 7.1  & 0.36 & 97.9 & 96.3 & 99.2 & 1841 \\
			& Hivecoin    & 12.1 & 0.35 & 97.9 & 97.1 & 99.7 & 1723 \\
			& FakeAV    & 13.2 & 0.59 & 97.1 & 96.2 & 98.6 & 1836 \\
			& Lokibot    & 14.7 & 0.57 & 97.1 & 96.5 & 99.4 & 1885 \\
			& Keylogger     & 20.2 & 0.57 & 98.4 & 97.1 & 99.2 & 2056 \\
			& Zeus    & 20.1 & 0.55 & 97.7 & 97.1 & 99.3 & 1861 \\
			& Ransom & 22.3 & 0.69 & 97.5 & 96.8 & 98.6 & 1880 \\
			& Rootkit    & 12.3 & 0.60 & 97.5 & 95.3 & 98.7 & 1869 \\
			\cmidrule{2-8}
			& \textit{Mean} & 15.3 & 0.54 & 97.6 & 96.55 & 99.09 & 1869 \\
			\midrule
			\multirow{9}{*}{BiLSTM}
			& Ramnit  & 6.6  & 0.53 & 98.2 & 97.7 & 99.5 & 2717 \\
			& Hivecoin    & 9.4  & 0.58 & 97.9 & 96.8 & 99.7 & 2816 \\
			& FakeAV    & 10.2 & 0.54 & 97.2 & 95.4 & 99.2 & 2761 \\
			& Lokibot    & 12.2 & 0.66 & 97.2 & 96.5 & 99.3 & 2837 \\
			& Keylogger     & 23.6 & 0.78 & 98.4 & 96.4 & 99.1 & 2685 \\
			& Zeus    & 25.0 & 0.89 & 97.9 & 97.2 & 99.5 & 2748 \\
			& Ransom & 16.8 & 0.62 & 97.7 & 97.0 & 98.6 & 2747 \\
			& Rootkit    & 13.7 & 0.81 & 97.1 & 94.9 & 99.1 & 2833 \\
			\cmidrule{2-8}
			& \textit{Mean} & 14.7 & 0.68 & 97.7 & 96.49 & 99.25 & 2768 \\
			\bottomrule
		\end{tabular}
	\end{adjustbox}
\end{table*}

	\textbf{Analysis.}
	XGBoost achieved the highest mean test accuracy across binary tasks (97.46\%) with a near-perfect mean AUC of 99.66\%, outperforming all other models including deep learning architectures. Random Forest followed closely (97.06\% accuracy, 99.52\% AUC) while consuming negligible RAM ($<$20\,MB). SVM, although fast to train and memory-lean, produced the lowest mean accuracy (94.88\%) and AUC (94.31\%), particularly struggling on the Ramnit family where its AUC fell to 87.76\%.
	
	Among deep learning models, LSTM achieved a slightly higher mean test accuracy (96.55\%) than BiLSTM (96.49\%), though BiLSTM produced a higher mean AUC (99.25\% vs.\ 99.09\%), reflecting better probabilistic calibration. Both recurrent models required 1.7–2.8\,GB peak RAM (roughly 40$\times$ or more than the classical models), with training times of 6–25 minutes per family, making them unsuitable for resource-constrained endpoints.
	
	These findings directly inform FDM input dimension~Q2 (\emph{Resource Budget Index}): for deployments with limited RAM or CPU, XGBoost provides the best accuracy-to-resource ratio while also attaining the highest AUC overall. Among the deep models, BiLSTM yields the best AUC (99.25\%); however, it does not surpass XGBoost on either accuracy or AUC and is justified only in resource-rich deployments that specifically require recurrent modelling of sequential features.

\subsubsection{Multi-Class Classification}

	In this experiment we extend the evaluation to simultaneous classification of all malware families (multi-class setting). The same sequence-splitting pre-processing and Optuna hyperparameter optimisation are applied. This represents the more realistic operational scenario where a deployed system must distinguish among all known families without separate binary detectors.

Tables~\ref{tab:exp1_multiclass_classic} and~\ref{tab:exp1_multiclass_dl} report results on the combined dataset (train: 17,795 / val: 5,932 / test: 5,932 sequences).

\begin{table}[htbp]
	\centering
	\caption{Multi-class API malware classification: classical ML models.}
	\label{tab:exp1_multiclass_classic}
	\begin{tabular}{lcccc}
		\toprule
		{Model} & {Val Acc} & {Test Acc} & {Train Time} & {RAM} \\
		& {(\%)}    & {(\%)}     & {(min)}       & {(MB)} \\
		\midrule
		XGBoost       & \textbf{79.74} & \textbf{79.03} & 353.9 & 189.4 \\
		Random Forest & 78.94 & 78.22 & 100.4 & 16.5  \\
		SVM           & 61.83 & 60.94 & 90.6  & 101.7 \\
		\bottomrule
	\end{tabular}
\end{table}

\begin{table}[htbp]
	\centering
	\caption{Multi-class API malware classification: deep learning models.}
	\label{tab:exp1_multiclass_dl}
	\begin{tabular}{lcccc}
		\toprule
		{Model} & {Val Acc} & {Test Acc} & {Train Time} & {RAM} \\
		& {(\%)}    & {(\%)}     & {(min)}       & {(MB)} \\
		\midrule
		BiLSTM & 75.29 & 72.27 & 62.1  & 2854.7 \\
		LSTM   & 73.60 & 73.18 & 72.1  & 2545.2 \\
		\bottomrule
	\end{tabular}
\end{table}

	\textbf{Analysis.}
	Multi-class classification reveals a marked drop in accuracy for all models compared to binary tasks. XGBoost retains the lead at 79.03\% test accuracy, while both LSTM and BiLSTM fall below 74\%, contrary to their strong binary performance. This reversal, in which classical models outperform deep sequence models in the multi-class case, is attributed to the greater structural diversity of API call patterns across nine simultaneous classes, which demands feature representations that sequence models have insufficient capacity to discriminate without significantly larger training corpora.
	
	For the FDM, this finding informs recommendation code~P2.1 (\emph{ML model selection}): in the multi-class setting, resource-efficient tree-based ensembles outperform deep sequence models, so the FDM should recommend classical classifiers unless the training corpus per class is substantially larger.

\subsection{Experiment~2: Class-Incremental Learning}
\label{subsec:exp2}

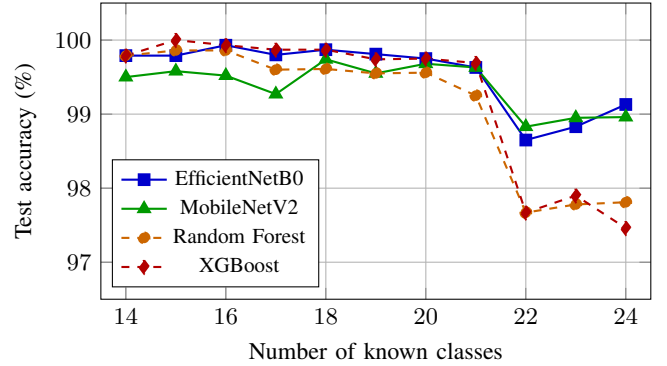
\begin{figure}[htbp]
	\centering
	\begin{tikzpicture}
		\begin{axis}[
			width=\columnwidth,
			height=5.5cm,
			xlabel={Number of known classes},
			ylabel={Test accuracy (\%)},
			xmin=13.5, xmax=24.5,
			ymin=96.5, ymax=100.5,
			xtick={14,16,18,20,22,24},
			ytick={97,98,99,100},
			legend style={at={(0.02,0.04)}, anchor=south west,
				font=\footnotesize, cells={align=left}},
			grid=both,
			grid style={line width=0.2pt, draw=gray!30},
			major grid style={line width=0.4pt, draw=gray!60},
			tick label style={font=\small},
			label style={font=\small},
			]
			\addplot[color=blue!80!black, mark=square*, mark size=2pt, thick]
			coordinates {
				(14,99.79)(15,99.79)(16,99.93)(17,99.80)(18,99.87)
				(19,99.81)(20,99.75)(21,99.63)(22,98.65)(23,98.83)(24,99.13)
			};
			\addlegendentry{EfficientNetB0}
			\addplot[color=green!60!black, mark=triangle*, mark size=2.5pt, thick]
			coordinates {
				(14,99.50)(15,99.58)(16,99.52)(17,99.27)(18,99.74)
				(19,99.55)(20,99.68)(21,99.63)(22,98.83)(23,98.95)(24,98.96)
			};
			\addlegendentry{MobileNetV2}
			\addplot[color=orange!80!black, mark=*, mark size=2pt, thick, dashed]
			coordinates {
				(14,99.79)(15,99.86)(16,99.86)(17,99.60)(18,99.61)
				(19,99.55)(20,99.56)(21,99.25)(22,97.67)(23,97.78)(24,97.81)
			};
			\addlegendentry{Random Forest}
			\addplot[color=red!70!black, mark=diamond*, mark size=2.5pt, thick, dashed]
			coordinates {
				(14,99.79)(15,100.0)(16,99.93)(17,99.87)(18,99.87)
				(19,99.74)(20,99.75)(21,99.69)(22,97.67)(23,97.90)(24,97.46)
			};
			\addlegendentry{XGBoost}
		\end{axis}
	\end{tikzpicture}
	\caption{Test accuracy of four models as new malware classes are introduced
		incrementally (Malimg dataset, classes 14–24). CNN-based models
		(EfficientNetB0, MobileNetV2) exhibit substantially less
		catastrophic forgetting than tree-based models.}
	\label{fig:exp2_cil_curve}
\end{figure}

%
	This experiment assesses how well models accommodate new malware families without full retraining, directly evaluating FDM component~P4 (\emph{Model Updating Techniques}) under the \emph{incremental} update mode. The Malimg dataset \cite{malimg} was used; images were pre-processed as $64{\times}64$ grayscale inputs. Models started with 14 known families and were incrementally updated one class at a time, yielding eleven evaluation points as the number of known classes grew from 14 to 24 (Fig.~\ref{fig:exp2_cil_curve}).
	
	\emph{Catastrophic forgetting} is quantified as the absolute accuracy drop between the best-ever accuracy observed before any new class was introduced and the final accuracy after all classes were added. Lower values indicate better knowledge retention.

\begin{table}[htbp]
	\centering
	\caption{Class-Incremental Learning (CIL) summary results on Malimg.
		Catastrophic Forgetting~(CF) is the absolute accuracy drop
		from the initial (pre-increment) accuracy to the final accuracy.}
	\label{tab:exp2_cil_summary}
	\begin{tabular}{lSSS}
		\toprule
		{Model} & {Final Test} & {Total Run-} & {CF} \\
		& {Acc. (\%)}  & {time (s)}   & {(pp)} \\
		\midrule
		EfficientNetB0 & \textbf{99.13} & 345.4  & 0.65 \\
		MobileNetV2    & 98.96 & 223.6  & \textbf{0.54}  \\
		Random Forest  & 97.81 & \textbf{95.7}   & 1.98  \\
		XGBoost        & 97.46 & 913.6  & 2.33  \\
		\bottomrule
	\end{tabular}
\end{table}

	\textbf{Analysis.}
	As shown in Table~\ref{tab:exp2_cil_summary} and Fig.~\ref{fig:exp2_cil_curve}, CNN-based image classifiers (EfficientNetB0 and MobileNetV2) exhibited substantially greater resistance to catastrophic forgetting (CF of 0.65\,pp and 0.54\,pp, respectively) than tree-based models (XGBoost: 2.33\,pp; RF: 1.98\,pp). Both CNN models maintained accuracy above 98.9\% after accommodating the full set of evaluated classes.
	
	Among classical models, Random Forest converged approximately $9.5\times$ faster than XGBoost (95.7\,s vs.\ 913.6\,s) while suffering only marginally lower forgetting, an important trade-off for latency-sensitive update pipelines.
	
	From the FDM perspective, this experiment validates recommendation code~P4 (\emph{Model Updating Techniques}): for environments that expect frequent addition of new malware families (high~Q4 \emph{Update Frequency Score}), CNN-based architectures on visualised binary representations are preferable, as they degrade gracefully under incremental update cycles. Resource-limited endpoints (low~Q2) may instead accept RF's slightly higher forgetting in exchange for dramatically faster update times.

\subsection{Experiment~3: Transfer Learning for Model Initialisation}
\label{subsec:exp3}

%
	This experiment investigates whether initialising CNN models from ImageNet pre-trained weights provides measurable advantages over training from scratch in terms of accuracy and training efficiency. Three heterogeneous datasets are used, namely (i)~the API call dataset (sequences converted to image-like representations), (ii)~the Android manifest permission dataset, and (iii)~the Malimg grayscale image dataset, to understand whether the benefit of transfer learning depends on the input modality. These findings directly inform FDM dimension Q4 (\emph{Update Frequency Score}) and the feature-type conditioning of recommendation code~P4 (MA1).

\begin{table*}[htbp]
	\centering
	\caption{Transfer learning vs.\ training from scratch: Malimg grayscale image dataset (25-class classification).}
	\label{tab:exp3_malimg}
	\begin{tabular}{llcccc}
		\toprule
		{Model} & {Training} & {Test Acc (\%)} & {Train Time (s)} & {Test Time (s)} & {Speed-up} \\
		\midrule
		\multirow{2}{*}{DenseNet121}
		& Pretrained & \textbf{95.88} & \textbf{628.5}  & 18.4 & \multirow{2}{*}{$1.51\times$} \\
		& Scratch    & 94.77          & 946.0           & 17.6 & \\
		\multirow{2}{*}{ResNet50}
		& Pretrained & 95.45 & \textbf{317.0}  & 17.4 & \multirow{2}{*}{$4.21\times$} \\
		& Scratch    & 95.56 & 1335.7          & 17.8 & \\
		\multirow{2}{*}{EfficientNetB0}
		& Pretrained & 91.49 & \textbf{347.4}  & 15.9 & \multirow{2}{*}{$1.90\times$} \\
		& Scratch    & 94.03 & 659.2           & 18.1 & \\
		\multirow{2}{*}{MobileNetV2}
		& Pretrained & \textbf{92.55} & \textbf{312.4}  & 17.2 & \multirow{2}{*}{$1.57\times$} \\
		& Scratch    & 91.60 & 489.6           & 17.3 & \\
		\midrule
		\textit{Mean (pretrained)} & & 93.84 & 401.3 & 17.2 & \multirow{2}{*}{$\bm{2.14\times}$} \\
		\textit{Mean (scratch)}    & & 93.99 & 857.6 & 17.7 & \\
		\bottomrule
	\end{tabular}
\end{table*}

\begin{table*}[htbp]
	\centering
	\caption{Transfer learning vs.\ training from scratch: Android manifest permission dataset (binary classification).}
	\label{tab:exp3_android}
	\begin{tabular}{llccc}
		\toprule
		{Model} & {Training} & {Test Acc (\%)} & {Train Time (s)} & {Test Time (s)} \\
		\midrule
		\multirow{2}{*}{DenseNet121}
		& Pretrained & \textbf{69.29} & 1535.5 & 38.0 \\
		& Scratch    & 36.99          & 1586.3 & 32.9 \\
		\multirow{2}{*}{ResNet50}
		& Pretrained & \textbf{67.19} & {—}    & {—}  \\
		& Scratch    & 67.19          & {—}    & {—}  \\
		\multirow{2}{*}{MobileNetV2}
		& Pretrained & 62.75 & \textbf{934.0}  & 32.9 \\
		& Scratch    & \textbf{63.01} & 819.2  & 39.4 \\
		\multirow{2}{*}{EfficientNetB0}
		& Pretrained & \textbf{68.36} & 1135.9 & 32.8 \\
		& Scratch    & 66.07          & 1261.6 & 33.3 \\
		\bottomrule
	\end{tabular}
\end{table*}

\begin{table*}[htbp]
	\centering
	\caption{Transfer learning vs.\ training from scratch: Windows API call dataset (binary classification; sequences rendered as 2-D feature maps).}
	\label{tab:exp3_api}
	\begin{tabular}{llccc}
		\toprule
		{Model} & {Training} & {Test Acc (\%)} & {Train Time (s)} & {Test Time (s)} \\
		\midrule
		\multirow{2}{*}{DenseNet121}
		& Pretrained & \textbf{69.65} & 1092.5 & 26.1 \\
		& Scratch    & 44.28          & 651.2  & 26.4 \\
		\multirow{2}{*}{ResNet50}
		& Pretrained & \textbf{61.19} & 469.7  & 15.6 \\
		& Scratch    & 40.80          & 436.3  & 15.4 \\
		\multirow{2}{*}{MobileNetV2}
		& Pretrained & 18.41 & \textbf{578.3}  & 16.6 \\
		& Scratch    & \textbf{30.85} & 738.1  & 16.4 \\
		\multirow{2}{*}{EfficientNetB0}
		& Pretrained & \textbf{72.64} & \textbf{456.1}  & 16.4 \\
		& Scratch    & 26.37          & 398.9  & 16.8 \\
		\bottomrule
	\end{tabular}
\end{table*}

	\textbf{Analysis.}
	Tables~\ref{tab:exp3_malimg}--\ref{tab:exp3_api} report results across all three datasets. Transfer learning offers a consistent training-time advantage on genuine image data (Malimg): pre-trained models complete training in an average of 401.3\,s vs.\ 857.6\,s for scratch-trained models ($2.14\times$ speed-up), while delivering comparable accuracy (93.84\% vs.\ 93.99\%). DenseNet121 and MobileNetV2 are the models whose pre-trained variant surpasses its scratch counterpart on both accuracy and training speed.
	
	For non-image modalities the picture is more nuanced. On the Android manifest permission dataset (Table~\ref{tab:exp3_android}), pre-trained CNN models generally outperform scratch-trained counterparts (DenseNet121: 69.29\% vs.\ 36.99\%), suggesting that low-level visual features learned from ImageNet still provide a useful inductive bias even for tabular–binary feature maps. On the raw API call dataset (Table~\ref{tab:exp3_api}), accuracy is substantially lower for all configurations (18–73\%), confirming that 2-D renderings of sequential API data do not form the natural images for which ImageNet representations were optimised.
	
	From an FDM standpoint, these results validate the following guidance encoded in recommendation~P4: (a)~pre-training accelerates convergence on image-based malware detection without sacrificing accuracy; (b)~for non-image feature types (e.g., API sequences), transfer learning from ImageNet provides limited benefit, and purpose-built architectures such as LSTM/BiLSTM are preferred; (c)~EfficientNetB0 and DenseNet121 are the most reliable pretrained backbones across datasets.

\subsection{Experiment~4: Autoencoder Feature Extraction}
\label{subsec:exp4}

%
	Feature dimensionality reduction is a practical concern for resource-limited endpoints. This experiment compares a two-stage pipeline (autoencoder pre-training followed by CNN classification on compressed features) against a single-stage CNN baseline. Both pipelines are evaluated on the Malimg dataset, with images resized to $64{\times}64$ pixels and flattened to 4,096-dimensional vectors (60\,\% train / 20\,\% validation / 20\,\% test, Optuna-optimised hyperparameters). This validates FDM recommendation code~P3.2 (\emph{Pre-processing}).

\begin{table*}[htbp]
	\centering
	\caption{Autoencoder feature extraction vs.\ CNN baseline on Malimg (25-class classification). \emph{Trainable Params} counts the parameters updated during supervised training; \emph{Total Params} counts all parameters (for the autoencoder pipeline, the 10.7\,M autoencoder plus the 3.4\,M classifier).}
	\label{tab:exp4_autoencoder}
	\begin{tabular}{lcccccc}
		\toprule
		{Pipeline} & {Test Acc.} & {Val Acc.} & {Train Time} & {Test Time} & {Trainable} & {Total} \\
		& {(\%)} & {(\%)} & {(min)} & {(s)} & {Params (M)} & {Params (M)} \\
		\midrule
		CNN baseline         & \textbf{96.21} & \textbf{96.47} & 117.7 & 8.51 & 27.80 & 27.80 \\
		Autoencoder + CNN    & 95.35          & 96.36          & \textbf{8.3} & \textbf{1.97} & \textbf{0.23} & \textbf{14.11} \\
		\midrule
		$\Delta$ (autoencoder vs. baseline) & $-0.86$ & $-0.11$ & $-109.4$ & $-6.54$ & $-27.57$ & $-13.69$ \\
		\bottomrule
	\end{tabular}
\end{table*}

	\textbf{Analysis.}
	As shown in Table~\ref{tab:exp4_autoencoder}, the autoencoder pipeline reduces training time by $14.2\times$ (from 117.7\,min to 8.3\,min) and test-time inference by $4.3\times$ (8.51\,s to 1.97\,s) in exchange for a modest accuracy penalty of 0.86\,pp on the test set and only 0.11\,pp on the validation set. The autoencoder encoder compresses 4,096-dimensional pixel vectors into a compact latent representation, so the supervised classifier head updates only 0.23\,M trainable parameters, versus 27.8\,M for the single-stage baseline CNN (a ${\sim}119\times$ reduction), which is the principal driver of the $14.2\times$ training speedup. The classifier model totals 3.4\,M parameters (an $8.1\times$ reduction in classifier size), and the full two-stage pipeline, including the 10.7\,M autoencoder, totals 14.1\,M parameters, about half the baseline. Equivalently, the autoencoder pipeline yields roughly $14\times$ more accuracy per training hour (693 vs.\ 49).
	
	These results confirm that autoencoder-based feature extraction is viable when training time or inference latency is a binding constraint, and the 0.86\,pp accuracy trade-off is operationally acceptable. This aligns with FDM guidance: organisations with a high \emph{Response Latency Level}~Q3 or low \emph{Resource Budget}~Q2 should consider autoencoder pre-processing (P3.2) to reduce classifier complexity without material accuracy loss.

\subsection{Framework Validation}\label{subsec:validation}

The cross-validation of the FDM recommendations against the empirical results in this sections is as follows. For a constrained, real-time deployment (Q1=1, Q2=1, Q3=3), the WCCS assigns the highest compatibility score to RF and XGBoost (RS=1). The binary classification experiments confirm this: XGBoost achieved 97.46\% mean test accuracy and RF 97.06\%, while consuming only 69\,MB and near-zero RAM respectively, compared to BiLSTM's peak of 2.8\,GB. For continuous-update deployments (Q4=3), the FDM recommends class-incremental learning (MA2); the CIL experiments confirm that EfficientNetB0 and MobileNetV2 achieve the lowest catastrophic-forgetting rates (0.65 and 0.54 percentage points respectively) and maintain above 98.9\% accuracy as the class count grows across the incremental steps (Fig.~\ref{fig:exp2_cil_curve}). For moderate-resource, image-feature deployments (Q2=2, FT1), the FDM recommends transfer learning (MA1); experiments confirm that pretrained models achieve equivalent accuracy to scratch training in an average of $2.14\times$ less time across four CNN architectures on the Malimg dataset.

\subsection{Key Findings}
\label{subsec:findings}

%
	Table~\ref{tab:fdm_validation} synthesises how each experimental outcome maps to the FDM input dimensions~(Q1–Q5) and recommendation codes~(P1–P5). Taken together, the four experiments demonstrate that no single ML configuration dominates across all operational contexts; this is precisely the condition the FDM is designed to address.
	
	\begin{enumerate}
		\item \textbf{Model selection is resource-sensitive (Q1, Q2 $\to$ P2):}
		Binary classification results confirm that XGBoost offers the best accuracy-resource balance for endpoints with moderate RAM budgets ($<$70\,MB) while also achieving the highest AUC overall; BiLSTM, the strongest deep model on AUC (99.25\%), still trails XGBoost and demands far more memory ($\sim$2.8\,GB RAM, $>$6\,min per family).
		
		\item \textbf{Multi-class complexity reverses model ranking (task complexity $\to$ P2.1):}
		The nearly 7\,pp accuracy gap between XGBoost (79.03\%) and BiLSTM (72.27\%) in multi-class classification shows that deep sequence models require substantially larger corpora to generalise across many families simultaneously, a finding that FDM encodes in the model-selection guidance of P2.1.
		
		\item \textbf{Incremental update strategy depends on update frequency and resource constraints (Q2, Q4 $\to$ P4):}
		EfficientNetB0 achieves the lowest catastrophic forgetting (0.65\,pp) but requires GPU-class resources; RF achieves comparable update speed ($\sim$96\,s) with higher forgetting (1.98\,pp) but negligible RAM, a trade-off explicitly parameterised in FDM via Q2 and Q4.
		
		\item \textbf{Transfer learning is modality-dependent (Q4, P3 $\to$ P4/MA1):}
		Pre-trained ImageNet weights provide a $2.14\times$ training speed-up on genuine malware images (Malimg) but negligible or negative benefit on API call–derived feature maps, confirming that FDM should condition transfer-learning recommendations on the selected feature type~P3.
		
		\item \textbf{Autoencoder pre-processing is preferable under latency or memory constraints (Q2, Q3 $\to$ P3.2):}
		A $14\times$ training acceleration with only 0.86\,pp accuracy cost positions autoencoder pipelines as the FDM-recommended pre-processing choice whenever Q3 is high (low response latency tolerance) or Q2 is low (tight resource budget).
	\end{enumerate}
	
	\begin{table*}[htbp]
		\centering
		\caption{Mapping of experimental findings to FDM input dimensions and
			recommendation codes. Q1~=~Platform Constraint, Q2~=~Resource Budget,
			Q3~=~Response Latency, Q4~=~Update Frequency, Q5~=~Sensitivity Ratio.}
		\label{tab:fdm_validation}
		\begin{adjustbox}{width=\textwidth}
			\begin{tabular}{p{3.5cm} p{2.5cm} p{2cm} p{5.5cm}}
				\toprule
				\textbf{Experiment / Finding} & \textbf{FDM Inputs} & \textbf{FDM Code} & \textbf{Validated Recommendation} \\
				\midrule
				Binary classification: XGBoost best accuracy-to-RAM ratio & Q1 (platform constraint), Q2 (low budget) & P2.1 & Prefer gradient-boosted tree over DL when RAM $<$100\,MB \\
				\addlinespace
				Binary classification: BiLSTM best AUC among DL models & Q1 (platform constraint), Q2 (high budget), Q3 & P2.1 & Use BiLSTM only in resource-rich settings; XGBoost still leads on accuracy and AUC \\
				\addlinespace
				Multi-class: classical models outperform DL & Task complexity (many classes) & P2.1 & Recommend tree ensemble for multi-class when data per class is limited \\
				\addlinespace
				CIL: CNN models resist catastrophic forgetting & Q4 (high update freq.) & P4 & Use CNN-based incremental learner for frequent new-family additions \\
				\addlinespace
				CIL: RF fastest update cycle & Q2 (low budget), Q4 & P4 & RF acceptable for low-resource incremental update pipelines \\
				\addlinespace
				Transfer learning: $2\times$ speed-up on image data & Q4 (update freq.) & P4 & Pre-train on ImageNet when feature type is malware visualisation \\
				\addlinespace
				Transfer learning: no benefit on API features & Q4, P3 choice & P4 & Avoid ImageNet transfer when feature type is API sequence \\
				\addlinespace
				Autoencoder: $14\times$ training speed-up, $-0.86$\,pp accuracy & Q2, Q3 & P3.2 & Deploy autoencoder pre-processing when latency or RAM is binding \\
				\addlinespace
				Sequence splitting (1,000-call segments): $>$94\% accuracy & P3.1 & P3.1 & Segment API traces at 1,000 calls; validated against raw truncation \\
				\bottomrule
			\end{tabular}
		\end{adjustbox}
	\end{table*}
	
	Collectively, these results demonstrate that the FDM's quantitative scoring mechanism (WCCS) correctly identifies the dominant performance-resource trade-off axes and steers practitioners toward configurations that are both high-performing and operationally feasible within their declared constraints. No individual model or pipeline achieves optimal performance across all five input dimensions simultaneously, underscoring the practical necessity of a structured decision-making framework.

\section{Discussion} \label{sec:discussion}

\subsection{Experimental Results}

The experimental results across all five tasks validate the FDM's core hypothesis: the optimal ML configuration is determined by operational constraints, not by accuracy alone.

\textbf{Resource and latency constraints (Q1, Q2, Q3).}
In binary API-based classification, XGBoost achieved the highest mean test accuracy of 97.46\% (AUC: 99.66\%) while consuming at most 69\,MB of RAM. In contrast, LSTM and BiLSTM achieved 96.55\% and 96.49\% mean accuracy respectively, but required 1.7--2.8\,GB peak RAM, a ${\sim}40\times$ overhead directly relevant to edge deployment. Random Forest matched deep learning accuracy (97.06\%) at near-zero RAM cost, reinforcing the FDM recommendation of classical ensemble models for low-Q2 deployments.

\textbf{Multi-class complexity and model ranking reversal (P2.1).}
A notable reversal emerged in multi-class classification: XGBoost achieved 79.03\% test accuracy while BiLSTM fell to 72.27\%, a gap of nearly 7\,pp. This directly contradicts the assumption that sequence-aware deep models are always superior for API-based tasks. The reversal arises because the current training corpus, split across nine simultaneous classes, provides insufficient per-class signal for recurrent models to generalise. This finding indicates that the FDM recommendation path for multi-class settings (many families, limited per-class data) should favour classical ensemble methods unless training data is substantially larger; hierarchical or ensemble deep architectures are identified as a direction for future work.

\textbf{Update frequency (Q4).}
Transfer learning on Malimg image data reduced average training time by $2.14\times$ (mean 401.3\,s pretrained vs.\ 857.6\,s scratch) with no statistically significant accuracy difference (93.84\% vs.\ 93.99\%). The benefit is strongest for ResNet50, which converged in $4.21\times$ fewer seconds (317\,s vs.\ 1,336\,s). However, on API-call feature maps and Android manifest vectors, pre-trained ImageNet weights provided limited advantage, confirming that the FDM should condition the MA1 (transfer learning) recommendation on the feature type (P3), not on the update interval alone. Class-incremental learning experiments support the MA2 (continuous-update) path: EfficientNetB0 and MobileNetV2 exhibited catastrophic forgetting of only 0.65 and 0.54 percentage points respectively across the incremental steps, while maintaining final accuracy above 98.9\%.

\textbf{Detection sensitivity and pre-processing efficiency (Q2, Q3, Q5).}
The autoencoder pipeline delivered a $14.2\times$ training speedup (117.7\,min $\to$ 8.3\,min) and a $4.3\times$ inference speedup (8.51\,s $\to$ 1.97\,s), at the cost of 0.86\,pp in test accuracy and 0.11\,pp in validation accuracy. This trade-off is well-characterised: when SR is low (FP-sensitive or latency-critical deployments), the speedup substantially outweighs the marginal accuracy loss; when SR is high (critical-infrastructure or low-tolerance environments), the full CNN pipeline remains warranted. This operating-point decision is directly encodable in FDM via Q3 and Q5.

\subsection{Limitations}

While the FDM provides a structured, mathematically grounded configuration methodology, six categories of deployment scenario fall outside its current scope.

\textbf{Federated and privacy-preserving deployments.}
The framework assumes training data can be centralised. Organisations subject to data-sovereignty regulations requiring federated learning (where only model updates, not raw samples, are exchanged) are not supported. A future input dimension (e.g., Data Locality Constraint) would address this.


\textbf{Adversarial and evasion-aware deployments.}
The framework does not account for malware crafted to evade ML detectors via feature-space perturbations or model inversion. Practitioners facing advanced persistent threats should treat P2.1--P2.3 as a baseline and layer adversarial hardening (e.g., adversarial training, certified defences) on top.



\textbf{Explainability requirements.}
FDM does not include explainability as an input dimension; P2.1 does not distinguish black-box models (e.g., CNNs) from interpretable ones (e.g., decision trees). Practitioners in regulated sectors (HIPAA, NIS2) must apply post-hoc methods such as SHAP or LIME, or manually override P2.1 recommendations.


\section{Conclusion}\label{sec:conclusion}

This paper presented the Framework for Decision-making (FDM) for building ML-based malware detection systems. The FDM formalises the configuration selection process through the Weighted Configuration Compatibility Score (WCCS), a multi-criteria scoring function that maps five quantifiable operational parameters, namely platform constraint (Q1), resource budget (Q2), response latency (Q3), update frequency (Q4), and detection sensitivity (Q5), to ranked recommendations across nine configuration dimensions covering data acquisition, model selection, hyperparameter optimisation, feature engineering, and model updating.

Empirical evaluation across five experimental tasks on three heterogeneous datasets (Windows API call sequences, Malimg grayscale images, and Android manifest permissions) validated the framework's recommendations. In binary malware classification, XGBoost achieved the highest accuracy-to-resource ratio (97.46\%, $<$70\,MB RAM), consuming approximately $40\times$ less memory than recurrent deep models while delivering competitive accuracy. Critically, in multi-class classification, this ranking reversed: XGBoost (79.03\%) outperformed BiLSTM (72.27\%) by nearly 7\,pp, confirming that the optimal model choice is task-complexity-dependent, a distinction the FDM explicitly encodes. For model update strategies, transfer learning reduced average training time by $2.14\times$ on image-based malware data with no accuracy cost (validating the MA1 periodic-update path), while class-incremental learning with EfficientNetB0 maintained 99.13\% final accuracy with only 0.65 percentage-point degradation across the incremental update steps (validating the MA2 continuous-update path). The autoencoder pre-processing trade-off (a $14\times$ training speedup against a 0.86\,pp accuracy loss) was quantified as a concrete operating point for latency- or resource-constrained deployments. Taken together, these results confirm the FDM's central premise: no single ML configuration dominates across all operational contexts, and structured, constraint-aware selection is both necessary and feasible.  

Future work will focus on two primary directions: (1) extending the WCCS input space to include dimensions for data-locality, explainability, and adversarial robustness; and (2) developing a federated and cross-platform variant of the recommendation algorithm.


\bibliographystyle{ieeetr}
\bibliography{ref}

\end{document}